\documentclass[sigconf]{acmart}
\usepackage{hyperref}
\usepackage{latexsym}
\usepackage{mathrsfs}
\usepackage{multirow}
\usepackage{amsmath}
\usepackage{amsfonts}
\usepackage{subfigure}
\usepackage{graphicx}
\usepackage{tabularx}
\usepackage[boxed,ruled,lined]{algorithm2e}
\usepackage{algorithmic}
\usepackage{comment}
\usepackage{balance}
\usepackage{bm}
\newtheorem{theorem}{Theorem}
\newtheorem{lemma}{Lemma}

\newcommand{\xc}[1]{{\color{blue} xuchen: ``#1''}}

\newcommand{\zhangx}[1]{{\color{green} zhangxiao: ``#1''}}

\AtBeginDocument{%
  \providecommand\BibTeX{{%
    \normalfont B\kern-0.5em{\scshape i\kern-0.25em b}\kern-0.8em\TeX}}}
\DeclareMathOperator*{\argmin}{arg\,min}
\DeclareMathOperator*{\argmax}{arg\,max}

\copyrightyear{2023}
\acmYear{2023}
\setcopyright{acmlicensed}\acmConference[WWW '23]{Proceedings of the ACM Web Conference 2023}{May 1--5, 2023}{Austin, TX, USA}
\acmBooktitle{Proceedings of the ACM Web Conference 2023 (WWW '23), May 1--5, 2023, Austin, TX, USA}
\acmPrice{15.00}
\acmDOI{10.1145/3543507.3583296}
\acmISBN{978-1-4503-9416-1/23/04}

\begin{document}
\begin{sloppy}
\title{P-MMF: Provider Max-min Fairness Re-ranking \\in Recommender System}
\author{Chen Xu}

\affiliation{%
  \institution{\mbox{Gaoling School of Artificial Intelligence}}
    \country{Renmin University of China}
  \\xc\_chen@ruc.edu.cn
}

\author{Sirui Chen}
\affiliation{%
  \institution{School of Information}
   \country{Renmin University of China}
  \\	csr16@ruc.edu.cn
}

\author{Jun Xu}
\authornote{Jun Xu is the corresponding author. Work partially done at Beijing Key Laboratory of Big Data Management and Analysis Methods.}
\affiliation{%
  \institution{\mbox{Gaoling School of Artificial Intelligence}}
    \country{Renmin University of China}
  \\junxu@ruc.edu.cn
}

\author{Weiran Shen}
\affiliation{%
  \institution{Gaoling School of Artificial Intelligence}
    \country{Renmin University of China}
  \\shenweiran@ruc.edu.cn
}

\author{Xiao Zhang}
\affiliation{%
  \institution{Gaoling School of Artificial Intelligence}
   \country{Renmin University of China}
  \\zhangx89@ruc.edu.cn
}

\author{Gang Wang\\Zhenghua Dong}
\affiliation{%
   \country{Huawei Noah's Ark Lab}
  \\	wanggang110@huawei.com
  \\dongzhenhua@huawei.com
}


\renewcommand{\shortauthors}{Chen Xu et al.} 
\begin{abstract}
In this paper, we address the issue of recommending fairly from the aspect of providers, which has become increasingly essential in multistakeholder recommender systems. 
Existing studies on provider fairness usually focused on designing proportion fairness (PF) metrics that first consider systematic fairness. However, sociological researches show that to make the market more stable, max-min fairness (MMF) is a better metric. The main reason is that MMF aims to improve the utility of the worst ones preferentially, guiding the system to support the providers in weak market positions. 
When applying MMF to recommender systems, how to balance user preferences and provider fairness in an online recommendation scenario is still a challenging problem. In this paper, we proposed an online re-ranking model named Provider Max-min Fairness Re-ranking (P-MMF) to tackle the problem. Specifically, P-MMF formulates provider fair recommendation as a resource allocation problem, where the exposure slots are considered the resources to be allocated to providers and the max-min fairness is used as the regularizer during the process. We show that the problem can be further represented as a regularized online optimizing problem and solved efficiently in its dual space. During the online re-ranking phase, a momentum gradient descent method is designed to conduct the dynamic re-ranking. Theoretical analysis showed that the regret of P-MMF can be bounded. Experimental results on four public recommender datasets demonstrated that P-MMF can outperformed the state-of-the-art baselines. Experimental results also show that P-MMF can retain small computationally costs on a corpus with the large number of items.

\end{abstract}



\maketitle
 
\section{Introduction}\label{sec:intro}


Raised out of social, ethical, and economic considerations, the fairness problem becomes non-negligible in recommendation~\cite{abdollahpouri2020multistakeholder,abdollahpouri2019multi}. In multi-stakeholder recommender systems (RS), there are several different participants, including users, items, providers, etc~\cite{abdollahpouri2020multistakeholder}. Various models have been proposed for recommending fairly from the viewpoints of users~\cite{li2021user}, providers~\cite{fairrec,fairrecplus}, or both~\cite{wang2021user}. In this paper, we are concerned about the problem of ensuring provider fairness in multi-stakeholder recommender systems.  

\begin{figure} 
    \centering    
    \includegraphics[width=0.8\linewidth]{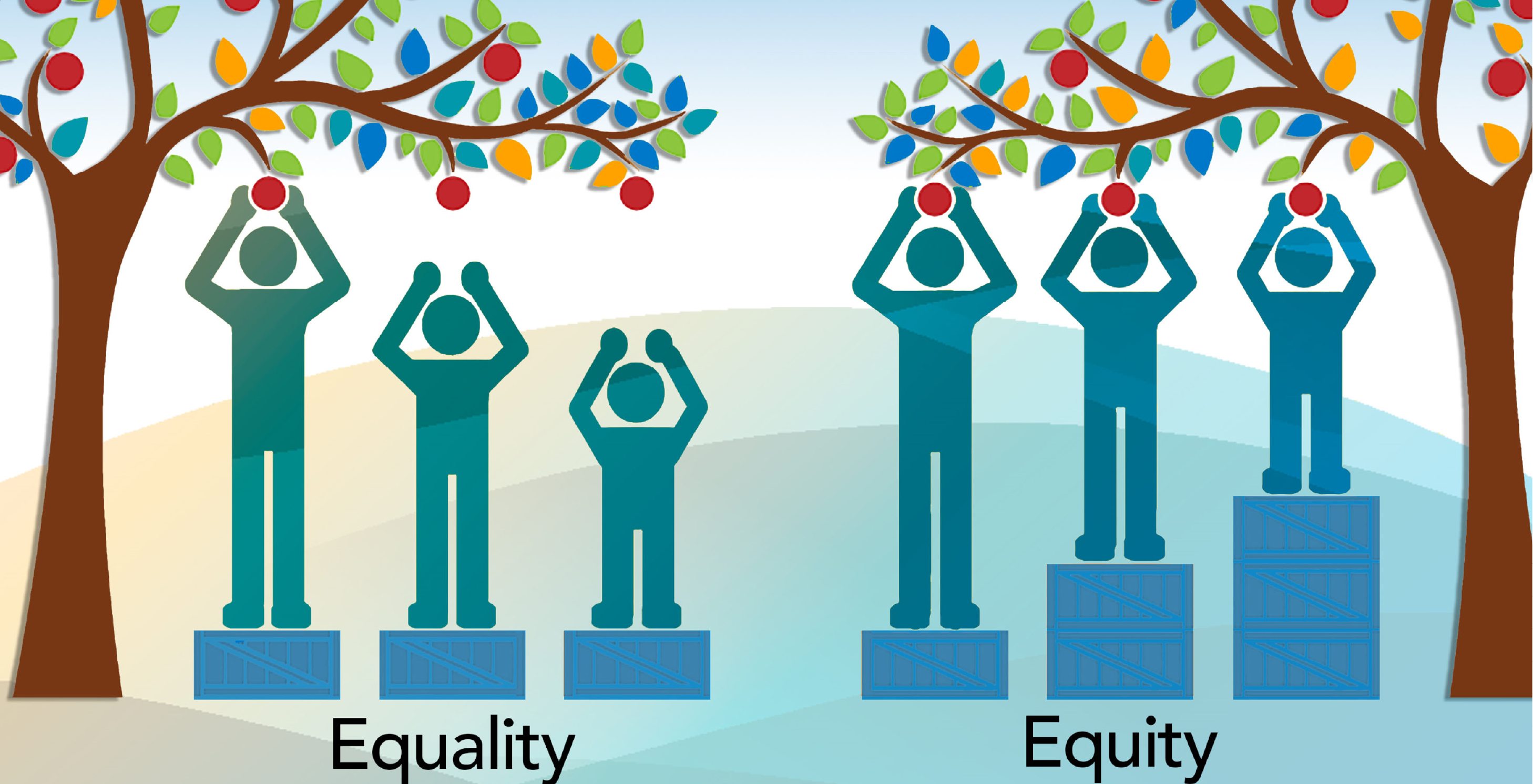}
    \caption{An intuitive example of equality and equity.\protect\footnotemark}
    \label{fig:intro}  
\end{figure}
\footnotetext{Image by Interaction Institute for Social Change | Artist: Angus Maguire. (interactioninstitute.org and madewithangus.com). Saskatoon Health Region.}

There are many cultural variations regarding fairness~\cite{tyler2002procedural,tyler1995social} in sociological researches. One practical fairness definition is based on two sociological concepts: equality and equity~\cite{matsumoto2016culture}. According to \citet{matsumoto2016culture}, equality can be defined as: \textit{everyone is treated the same and provided the same resources to succeed}, while equity can be defined as: \textit{ensuring that resources are equally distributed based on needs}. 
Figure~\ref{fig:intro} gives an intuitive example of the two types of fairness under the recommendation scenario. Suppose we have some resources (i.e., exposure slots) to ensure that providers can get the apples (i.e., survival in the market). As shown in Figure~\ref{fig:intro}, equality ensures that the RS evenly gives each provider the same support, while equity (known as distributive justice) emphasizes that RS will assign resources to providers as different ratios~\cite{langemeyer2020weaving}.


Based on the concepts of equity~\cite{lamont2017distributive}, the metrics of proportion fairness (PF) and max-min fairness (MMF)~\cite{nash1950bargaining} have been respectively proposed. PF and MMF have been widely used for computation network and transportation~\cite{peterson2007computer,dondeti1996max,bonald2006queueing}. Their formal formulations in the provider fair recommendation scenario will be explained in Section~\ref{sec:formulation}. Intuitively, PF and MMF tries to assign resources to the specified ratios~\cite{price4fairness}. PF is based on welfare-based principles (known as Aristotle’s Principle and Nash solution~\cite{nash1950bargaining}), which maximizes the sum of welfare of providers. MMF (known as Rawls’ Principle and Kalai-Smorodinsky solution~\cite{kalai1975other}) aims to improve the worst-off providers’ utilities. MMF has proven to be a better metric for the provider fairness problem, because worse-off providers, who occupy the majority of the platform, can survive with these supports. Taking care of these weak providers will increase the stability of the recommender market~\cite{fairrec}. For example, according to the report by \citet{Erickson18}, small sellers in Amazon have difficulty facing the challenges of ``lower profitability'' and ``inability to personalize'' on their own. The unfair system without ensuring MMF may result in a broken relationship between the providers and RS and, finally, force the providers to leave.


Existing provider fair recommendation models either consider the PF of providers~\cite{cpfair,fairrec,fairrecplus,wu2021tfrom} or use heuristics to guarantee the utilities of the worse-off providers~\cite{fairrec,fairrecplus}. However, these heuristic methods do not directly consider the MMF and lack theoretical guarantees when adapting to online scenarios. Moreover, the heuristic methods lack the flexibility to trade-off with user utility, which inevitably hurts the users' experience.  



This paper aims to develop a practical re-ranking model that considers MMF from the providers' perspectives.
We formulate the provider fair recommendation as a process of resource allocation~\cite{bower1972managing}. In such a process, resources can be regarded as limited ranking slots, and providers are viewed as the demanders. The cost of allocation is defined as the preference of the users. Moreover, an MMF regularizer is imposed on the allocation to maximize the minimum allocation to a specific provider. 


To adapt the method to the online recommendation scenarios, we proposed an online re-ranking algorithm called P-MMF that focuses on provider max-min fairness by viewing the provider fair online recommendation as a regularized online optimizing problem. However, the optimization problem contains many integral variables, making it notoriously difficult to solve. We then derive its dual problem and propose an efficient algorithm to optimize the problem in the dual space. In the online setting, a momentum gradient descent method was developed to make an effective and efficient online recommendation. Our theoretical analysis shows that the regret of the P-MMF can be well-bounded. Furthermore, the P-MMF is also computationally efficient due to its insensitivity to the number of items.


We summarize the major contributions of this paper as follows:

(1) We analyzed the importance of ensuring equity for providers in multi-stakeholder recommender systems, and propose to use the max-min fairness metric in provider fair recommendation.

(2) We formulated the provider fair recommendation as a resource allocation problem regularized by the max-min fairness, and proposed a re-ranking model called P-MMF that balances the provider fairness and the user preference. Our theoretical analysis showed that the regret of P-MMF can be bounded.

(3) Simulations on a small dataset showed the superiority of MMF compared to PF in recommending fairly to providers. Extensive experiments on four publicly available datasets demonstrated that P-MMF outperformed state-of-the-art baselines, including the PF-based and MMF-based methods. 

\section{Related Work}
Fairness has become a hot research topic in multi-stakeholder recommender systems~\cite{abdollahpouri2020multistakeholder,abdollahpouri2019multi}. Researchers have proposed several user-oriented and item/provider-oriented fairness re-ranking models. For user-oriented fairness,~\citet{abdollahpouri2019unfairness} proposed a user-centered evaluation that measures users' different interest levels in popular items.~\citet{li2021user,serbos2017fairness,sacharidis2019top} addressed the user-centered fairness from a group fairness perspective. For provider-oriented fairness,~\citet{gomez2022provider} proposed a rule-based algorithm to ensure that the exposures of items should be equally distributed.~\citet{ge2021towards} aimed to enhance dynamic fairness when item popularity changes over time. However, these provider-oriented methods did not consider the user's perspective.

Recently, there are also studies that jointly consider the trade-off between user preference and provider fairness.~\citet{chakraborty2017fair} claimed that all providers should receive the amount of exposures proportional to their relevance in economy platforms.~\citet{rahmani2022unfairness} studied the trade-offs between the user and producer fairness in Point-of-Interest recommendations. TFROM~\cite{wu2021tfrom} and CPFair~\cite{cpfair} formulated the trade-off as a knapsack problem and a relaxed linear programming problem, respectively. However, they used greedy-based algorithms in online scenarios, which only improves the proportion fairness of providers. Some studies noticed that the utilities of worse-off providers should also be guaranteed. For example, FairRec~\cite{fairrec} and its extension FairRec+~\cite{fairrecplus} proposed hard constraints to ensure that every provider should have the lowest exposures. Welf~\cite{nips21Welf} proposed a Frank-Wolfe algorithm to maximize the welfare functions of worse-off items. However, they were all designed for offline scenarios and suffered from high computational costs, which prevented them from being applied to real online recommendation systems \cite{Zhang2021Counterfactual,Zhang2022Counteracting}.

In this paper, we formulate the re-ranking task as the resource allocation problem~\cite{bower1972managing}, which is crucial in communications and transportation.  
In online resource allocation, most studies~\cite{devanur2011near,li2021online} focused on designing the time-separable reward functions, which is the sum of rewards over periods. ~\citet{li2020simple} proposed a local-based sub-gradient algorithm for the linear reward.~\citet{cheung2020online} designed a dual-based online algorithm with learning from the distribution of requests.~\citet{balseiro2021regularized} proposed a mirror-descent method to solve the fairness-regularized online allocation problem. 


\section{Provider Fair Re-ranking}\label{sec:formulation}
In this section, we first define the notations in a multi-stakeholders recommender system. Then we give the formal definitions of the proportion and max-min fairness. 

In a multi-stakeholders recommender system, multiple participants exist, including users, item providers, and other stakeholders. Let $\mathcal{U}, \mathcal{I}$, and $\mathcal{P}$ be the set of users, items, and providers, respectively. Each item $i\in \mathcal{I}$ is associated to a unique provider $p\in \mathcal{P}$. The set of items associated with a specific provider $p$ is denoted as $\mathcal{I}_p$. When a specific user $u\in \mathcal{U}$ accesses the recommender system, a list of $K$ items, denoted by $L_K(u)\in\mathcal{I}^K$, is provided to the user. For each user-item pair $(u, i)$, the recommender model estimates a preference score $s_{u,i}\in\mathbb{R}$. These items are ranked according to their preference scores. 
In this paper, we define the user-side utility of exposing item list $L_K(u)$ to $u$ as the summation of the preference scores in the list, denoted by $f\left(L_K(u)\right) = \sum_{i\in L_K(u)} s_{u,i}$. We follow the literature convention \cite{fairrec,cpfair} and define the fairness vector of providers as $\mathbf{e}\in\mathbb{R}^{|\mathcal{P}|}$, where for a specific provider $p$, $\mathbf{e}_p\in\mathbb{R}^+$ denotes the number of exposed items of provider $p$. The goal of provider fair re-ranking is to compute a new fair list $L^F_K(u)\in\mathcal{I}^K$ which well balances the user utilities $f\left(L^F_K(u)\right)$ and a provider fairness metric defined over $\mathbf{e}$. 


In real-world applications, the users arrive at the recommender system sequentially. Assume that at time $t$ user $u_t$ arrives. The recommender system needs to consider long-term provider exposure during the entire time horizon from $t=0$ to $T$.
Our task can be formulated as a resource allocation problem~\cite{balseiro2021regularized} with time-separable fairness. 
Specifically, the optimal utility of the recommender system can be defined as a time-separable utility function, which is the accumulated reward \cite{balseiro2021regularized} over periods from 0 to $T$. In this case, $\mathbf{e}_p$ can be seen as the total number of exposed items of provider $p$, accumulated over the period $0$ to  $T$. Formally, when trading-off the user preference and provider fairness,  we have the following mathematical program:
\begin{equation}
\label{eq:OPT}
\begin{aligned}
        \max_{L^F_K}&\quad \frac{1}{T}\sum_{t=1}^T f\left(L^F_K(u_t)\right) + \lambda r(\mathbf{e})\\
        \mathrm{s.t.} &\quad \mathbf{e} \leq \bm{\gamma}
\end{aligned},
\end{equation}
where $\bm{\gamma}\in\mathbb{R}^{|\mathcal{P}|}$ denotes the weights of different providers, e.g., weighted PF or MMF~\cite{price4fairness}, and $r(\mathbf{e})\in\mathbb{R}$ is a provider fairness metric that serves as a fairness regularizer. Note that the constraint $\mathbf{e}\leq \bm{\gamma}$ can also be viewed as the maximum resources allocated to the providers. Following the definitions of PF and MMF~\cite{price4fairness}, $r(\mathbf{e})$ also has two different forms:

\textbf{Proportion Fairness (PF)}: $r(\mathbf{e}) = \sum_{p\in \mathcal{P}} \log \left[1+\mathbf{e}_p/\bm{\gamma}_p\right]$.

\textbf{Max-min Fairness (MMF)}: $r(\mathbf{e}) = \min_{p\in \mathcal{P}} \left[\mathbf{e}_p/\bm{\gamma}_p\right]$.

Using the proportion fairness as the regularizer, we can reduce the difference between $\mathbf{e}$ and $\bm{\gamma}$, while using the max-min fairness we can improve the relative exposure (w.r.t. weights $\bm{\gamma}$) of the least exposed provider.

\section{Our approach: P-MMF}
In this section, we first formulate provider fairness in the recommendation as a resource allocation problem. Then, we propose an algorithm called P-MMF for the online recommendation.

\subsection{Resource allocation with provider fairness}
We formulate the providers' fair recommendation problem as a resource allocation process~\cite{bower1972managing}. In such a process, the resources can be regarded as limited ranking slots that are allocated to providers. The allocation cost is defined based on the preference of the users and the max-min fairness. 

Formally, based on Equation~\eqref{eq:OPT}, the fair recommendation problem can be written as a linear programming:
\begin{equation}
\label{eq:solve_OPT}
\begin{aligned}
         \max_{\mathbf{x}_t} \quad& \frac{1}{T}\sum_{t=1}^Tg(\mathbf{x}_t) + \lambda r(\mathbf{e})\\
        \textrm{s.t.}\quad  
        &\sum_{i\in\mathcal{I}}\mathbf{x}_{ti} = K, 
        \quad \forall t\in [1,2,\ldots,T]\\
         &\mathbf{e}_p = \sum_{t=1}^T\sum_{i\in\mathcal{I}_p}\mathbf{x}_{ti}, 
         \quad \forall p\in \mathcal{P}\\
         &g(\mathbf{x}_t) = \sum_{i\in \mathcal{I}}\mathbf{x}_{ti}s_{u_t,i}, \quad \forall t\in [1,2,\ldots,T]\\
        &\mathbf{e} \leq \bm{\gamma}, \mathbf{x}_{ti} \in \{0, 1\}, 
        \quad \forall i \in\mathcal{I}, ~\forall t\in [1,2,\ldots,T]
\end{aligned},
\end{equation}
where $\mathbf{x}_t\in\{0, 1\}^{|\mathcal{I}|}$ is the decision vector for user $u_t$, and $g(\cdot)$ is the user-side utility function. Specifically, for each item $i$, $\mathbf{x}_{ti} = 1$ if it is added to the fair ranking list $L_K^F(u_t)$, otherwise, $\mathbf{x}_{ti} = 0$. Note that $g(\cdot)$ is equivalent to $f(\cdot)$ in the sense that they produce the same result, while $g(\cdot)$ takes the binary decision vector as input. The first constraint in Equation~\eqref{eq:solve_OPT} ensures that the recommended lists are of size $K$. The second constraint in Equation~\eqref{eq:solve_OPT} suggests that the exposures of each provider $p$ are the accumulated exposures of the corresponding items over all periods. In general, we think time-separable fairness would be preferred under the scenarios with weak timeliness. For example, recommending the items with long service life (e.g., games and clothes etc.).


\begin{figure}
    \centering
    \includegraphics[width=0.8\linewidth]{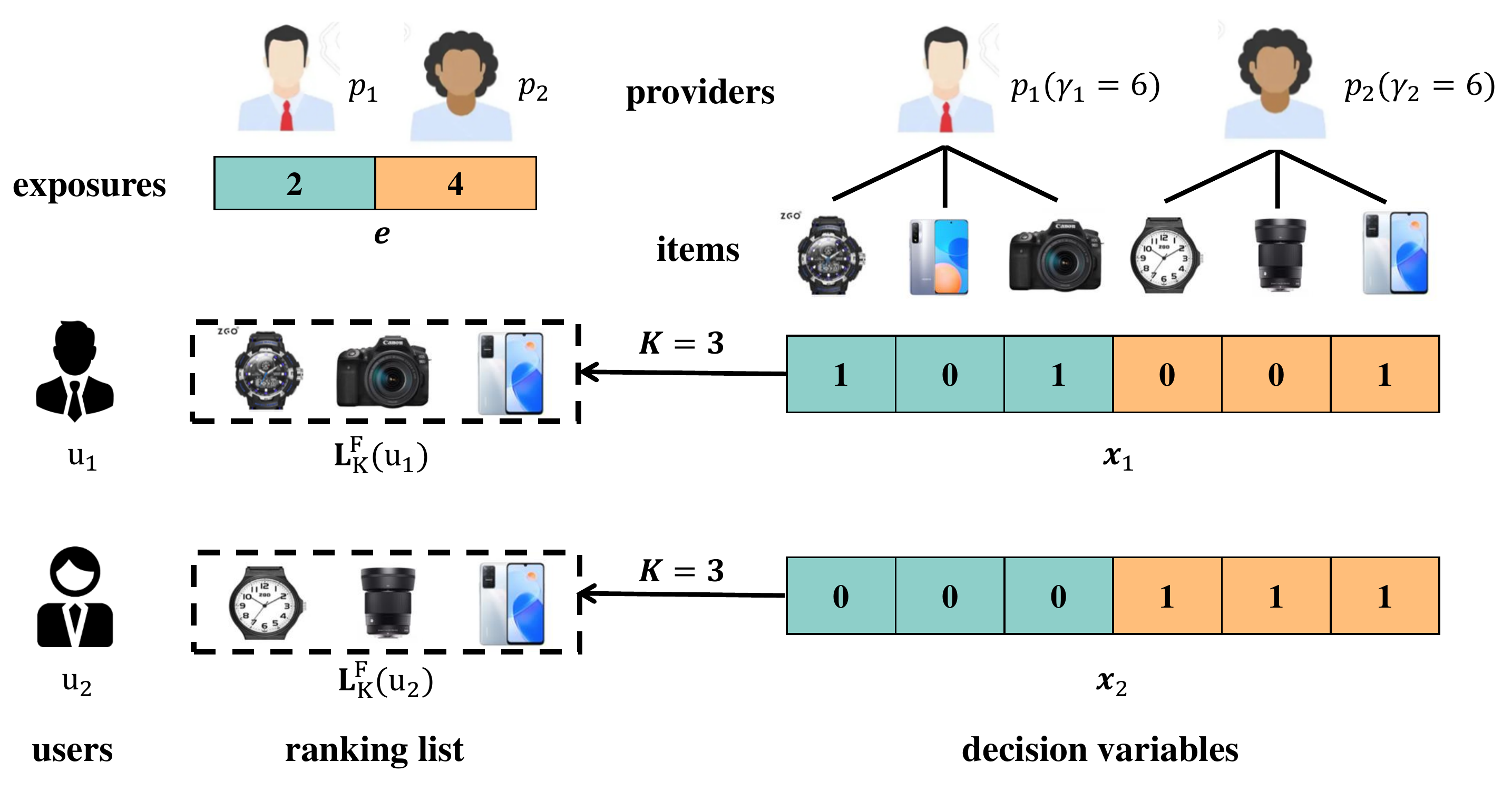}
    \caption{A toy example for time-separable fairness in recommendation}
    \label{fig:toy-example}
    \vspace{-0.6cm}
\end{figure}

Figure~\ref{fig:toy-example} gives a toy example of the linear programming problem~\eqref{eq:solve_OPT}. In the example, we set $T=2$. Suppose there are two users, $u_1$ and $u_2$, arriving at the system one by one. At each time step, the system recommends a list of $K=3$ items. Therefore, the system has overall $2\times 3 = 6$ slots to expose. Suppose the system has two providers $p_1$ and $p_2$, each owning three items. Let's set identical weights for $p_1$ and $p_2$ (i.e., $\bm{\gamma} = [\boldsymbol{\gamma}_1=6, \boldsymbol{\gamma}_2=6])$. The model solves problem~\eqref{eq:solve_OPT} and the solution is two binary vectors: $\mathbf{x}_1$ for $u_1$ and $\mathbf{x}_2$ for $u_2$. Finally, we count the exposures over the time $T=2$: $\mathbf{e} = [e_{p_1} = 2, e_{p_2} =4]$, which means by recommending the ranking lists ${L}_K^F(u_1)$ (created based on $\mathbf{x}_1$) and ${L}_K^F(u_2)$ (created based on $\mathbf{x}_2$) to $u_1$ and $u_2$, provider $p_1$ and $_2$ get 2 and 4 exposures, respectively.

Although here we have already given a linear programming solution Eq.\eqref{eq:solve_OPT} to the problem, it can only be solved small-scale problems in an offline way. In online recommendation systems, for each user $u_t$ access, the model needs to generate a fair ranking list $L^F_K(u_t)$ from large-scale item corpus immediately. This means we have no idea about the information after $t$. Next, we will discuss how to use MMF in the online recommendation problem.





\subsection{P-MMF for online applications}\label{sec:P-MMF}
In this section, firstly, we formulate the MMF in online scenarios. Next, we consider the dual of the original problem. Focusing on the dual problem has several advantages: the dual problem has significantly fewer variables, and the variables no longer need to be integers as in the original problem. Finally, we proposed a momentum gradient descent algorithm for efficient online learning. 

\subsubsection{\mbox{Max-min fairness in online scenarios}}
At time step $t$, the recommender system receives a request from user $u_t$. An online algorithm $h$ produces a real-time decision vector $\mathbf{x}_t\in \{0,1\}^{|\mathcal{I}|}$ based on the current user $u_t$ and the previous history
$\mathcal{H}_{t-1} = \{u_s,\mathbf{x}_s\}_{s=1}^{t-1}$:
$
    \mathbf{x}_t = h(u_t \mid \mathcal{H}_{t-1}).
$
We define the online reward of max-min fairness to be the summation of rewards over all time steps:
\begin{equation}\label{eq:W_eq}
    W = \frac{1}{T}\sum_{t=1}^Tg(\mathbf{x}_t) + \lambda \min\left[\mathbf{e}/\bm{\gamma}\right],
\end{equation}
where $\min\left[\mathbf{e}/\bm{\gamma}\right]$ corresponds to the max-min fairness regularizer. 

Our goal is to design an algorithm $h$ that attains low
regret. Denote by $W_{OPT}$ the optimal value, we measure the
regret of an algorithm as the expectation difference between the optimal performance $W_{OPT}$ of the offline problem and that of the online algorithm $W$ over user distributions $\mathcal{U}$:
\begin{equation}
    \text{Regret}(h) = \mathbb{E}_{u_t \sim \mathcal{U}} \left[W_{OPT} - W\right].
\end{equation}

\subsubsection{Dual problem}
From the original problem in Equation~\eqref{eq:OPT}, we know that the integer decision variable $\mathbf{x}_t$ is of size $|\mathcal{I}|$ for each time $t$, which is hard to solve. However, considering its dual problem,  we can significantly reduce its computational cost.

\begin{theorem}[Dual Problem]\label{theo:dual}
The dual problem of Equation~\eqref{eq:OPT} can be written as:
\begin{equation}
\label{eq:dual}
\begin{aligned}
        W_{OPT} &\leq W_{Dual} = \min_{\boldsymbol{\mu}\in\mathcal{D}}\left[g^*(\mathbf{A}\boldsymbol{\mu}) + \lambda r^*(-\boldsymbol{\mu})\right],
\end{aligned}
\end{equation}
where $\mathbf{A}\in\mathbb{R}^{|\mathcal{I}|\times|\mathcal{P}|}$ is the item-provider adjacent matrix, and $A_{ip} = 1$ indicates item $i\in \mathcal{I}_p$, and 0 otherwise.
Letting $\mathcal{X} = \{\mathbf{x}_t|\mathbf{x}_t \in \{0,1\} \land \sum_{i\in\mathcal{I}} \mathbf{x}_{ti} = K\}$, $g^*(\cdot),r^*(\cdot)$ are the conjugate functions:
\begin{align*}
    g^*(c) = \max_{\mathbf{x}_t\in\mathcal{X}}\sum_{t=1}^T\left[g(\mathbf{x}_t)/T - \mathbf{c}^\top\mathbf{x}_t\right], 
    r^*(-\boldsymbol{\mu}) = \max_{\mathbf{e}\leq \bm{\gamma}}\left[r(\mathbf{e})+\boldsymbol{\mu}^\top\mathbf{e}/\lambda\right],
\end{align*}
and $\mathcal{D} = \{\boldsymbol{\mu}|r^*(-\boldsymbol{\mu})<\infty\}$ is the feasible region of dual variable $\boldsymbol{\mu}$. 
\end{theorem}


Proof of Theorem~\ref{theo:dual} can be found in Appendix~\ref{app:dual_prove}. From Theorem~\ref{theo:dual}, we can have a new non-integral decision variable $\boldsymbol{\mu} \in \mathbb{R}^{|\mathcal{P}|}$. In practice, the provider size $|\mathcal{P}|\ll |\mathcal{I}|$. Besides, due to $\mathbf{A}$'s sparsity, it is very efficient to compute $\mathbf{A}\boldsymbol{\mu}$, which aims to project the variable $\boldsymbol{\mu}$ from provider space into item spaces.

In our online algorithm discussed in Section~\ref{sec:alg}, we can have a closed form of the conjugate function $g^*(\cdot)$ in constant time. As for the feasible region $\mathcal{D}$ and the conjugate function $r^*(\cdot)$, we have Theorem~\ref{theo:region} and Lemma~\ref{lem:reg_form}.

\begin{theorem}[Dual Feasible Region]\label{theo:region}
In the MMF, the feasible region of the dual problem 
\[
    \mathcal{D} = \left\{\boldsymbol{\mu} ~~\left|~~ \sum_{p\in\mathcal{S}} \bm{\gamma}_p\boldsymbol{\mu}_p \ge -\lambda, \forall \mathcal{S}\in\mathcal{P}_s\right.\right\},
\]
where $\mathcal{P}_s$ is the power set of $\mathcal{P}$, i.e., the set of all the subsets of $\mathcal{P}$.
\end{theorem}

The proof of Theorem~\ref{theo:region} can be found in Appendix~\ref{app:dual_prove}, which implies the following Lemma~\ref{lem:reg_form}:

\begin{lemma}\label{lem:reg_form}
The conjugate function $r^*(\cdot)$ has a closed form:
$
    \max_{\boldsymbol{\mu}\leq\boldsymbol{\gamma}}r^{*}(-\boldsymbol{\mu}) = \bm{\gamma}^T\boldsymbol{\mu}/\lambda + 1,
$
and the optimal dual variable is:
$
    \argmax_{\boldsymbol{\mu}\leq\boldsymbol{\gamma}}r^{*}(-\boldsymbol{\mu}) = \bm{\gamma}/\lambda.
$
\end{lemma}

\subsubsection{The P-MMF algorithm}\label{sec:alg}
Algorithm~\ref{alg:P-MMF} illustrates P-MMF algorithm. Following \citet{balseiro2021regularized}, P-MMF keeps a dual variable $\boldsymbol{\mu}_t$, the remaining resources $\boldsymbol{\beta}_t$ and the gradient $\mathbf{g}_t$ for each time $t$.

Whenever a user arrives, the algorithm computes the recommended variable $\mathbf{x}_t$ based on the remaining resources and the dual variable $\boldsymbol{\mu}_t$ (line 7). The final dual variable is estimated as the average dual variable for each time $t$: $\boldsymbol{\mu} = \sum_{t=1}^T\boldsymbol{\mu}_t/T$. Intuitively, for $\boldsymbol{\mu}_t$, when the values of dual variables are higher, the algorithm naturally recommends fewer items related to the corresponding provider. The remaining resources $\boldsymbol{\beta}_t$ ensure that the algorithm only recommends items from providers with remaining resources. Note that in line 7, the formulation is linear with respect to $\mathbf{x}_t$. Therefore, it is efficient to compute $\mathbf{x}_t$ through a top-$K$ sort algorithm in constant time.

The online learning process is as follows. Firstly, we get the closed form of the conjugate function of max-min regularizer $r^*(-\boldsymbol{\mu}_t)$ according to Lemma~\ref{lem:reg_form}. Then we can get the subgradient of the dual function $g^*(\mathbf{A}\boldsymbol{\mu})+\lambda r^*(-\boldsymbol{\mu})$:
\[
    -\mathbf{A}^{\top}\mathbf{x}_t + \mathbf{e}_t \in \partial \left(g^*(\mathbf{A}\boldsymbol{\mu}_t)+\lambda r^*(-\boldsymbol{\mu}_t)\right).
\]

We also add the last time momentum~\cite{qian1999momentum} to the updated gradient $\mathbf{g}_t$. Finally, we utilize $\mathbf{g}_t$ to update the dual variable by performing the online descent in line 14, where we used weighted $\ell_2-norm:\|\boldsymbol{\mu}\|_{\boldsymbol{\gamma}^2}^2 = \sum_{j=1}^{|\mathcal{P}|}\boldsymbol{\gamma}_j^2\boldsymbol{\mu}_j^2$. Therefore, the dual variable will move towards the directions of the providers with fewer exposures, and the primal variable $\mathbf{x}_t$ will move to a better solution.

Note that the projection step in line 14 can be efficiently solved using convex optimization solvers~\cite{balseiro2021regularized} since $\mathcal{D}$ is coordinate-wisely symmetric. 
\begin{align*}
    \min_{\boldsymbol{\mu}\in\mathcal{D}}\| \boldsymbol{\mu}-\boldsymbol{\mu}_t \|_{\boldsymbol{\gamma}^2}^2  &= \min_{\sum_{p\in\mathcal{P}}}(\boldsymbol{\mu}_p\boldsymbol{\gamma}_p - \widetilde{\boldsymbol{\mu}}_p\boldsymbol{\gamma}_p)^2\\
    \textrm{s.t.} \sum_{j=1}^m \boldsymbol{\gamma}_j\widetilde{\boldsymbol{\mu}}_j + \lambda &\ge 0,~ \forall m = 1, 2 , \ldots, |\mathcal{P}|,
\end{align*}
where $\widetilde{\boldsymbol{\mu}}$ satisfies:
$
    \boldsymbol{\gamma}_1\widetilde{\boldsymbol{\mu}}_1 \leq \boldsymbol{\gamma}_2\widetilde{\boldsymbol{\mu}}_2 \leq \cdots, \leq \boldsymbol{\gamma}_{|\mathcal{P}|}\widetilde{\boldsymbol{\mu}}_{|\mathcal{P}|}.
$

We provide a regret bound on $\text{Regret}(h)$ in Theorem~\ref{theo:regret} but defer the proof to Appendix~\ref{app:regret_prove} due to space limit. Intuitively, larger ranking size $K$ and time $T$ will lead to larger biases in P-MMF.




\begin{theorem}[Regret Bound]\label{theo:regret}
Assume that the function $\|\cdot\|_{\boldsymbol{\gamma}^2}^2$ is $\sigma$-strong convex and there exists a constant $G\in\mathbb{R}^{+}$ and $H>0$ such that $\|\widetilde{g}_t\|<G$, $ \| \boldsymbol{\mu}_t-\boldsymbol{\mu}_0\|_{\boldsymbol{\gamma}^2}^2\leq H$..
Then, the regret can be bounded as follows:
\begin{equation}
    \text{Regret}(h) \leq \frac{K(1+\lambda \bar{r} + \bar{r})}{\min_p \boldsymbol{\gamma}_p} + \frac{H}{\eta} + \frac{G^2}{(1-\alpha)\sigma}\eta(T-1) + \frac{G^2}{2(1-\alpha)^2\sigma\eta},
\end{equation}
where $\bar{r}$ is the upper bound of MMF regularzier, and in practice, $\bar{r}\leq 1$.
\end{theorem}
Setting the learning rate as $\eta  = O(T^{-1/2})$, we can obtain a sublinear regret upper bound of order  $O(T^{1/2})$ of magnitude.


\begin{algorithm}[t]
    \caption{Online learning of P-MMF}
	\label{alg:P-MMF}
	\begin{algorithmic}[1]
	\REQUIRE User arriving order $\{u_i\}_{i=1}^N$, time-separate size $T$, ranking size $K$, user-item preference score $s_{u,i}, \forall u,i$, item-provider adjacent matrix $\mathbf{A}$, maximum resources $\bm{\gamma}$ and the trade-off coefficient $\lambda$.
	\ENSURE The decision variables $\{\mathbf{x}_i, i = 1,2,\ldots, N\}$
	\FOR{$n=1,\cdots,N/T$}
	    \STATE Initialize dual solution $\boldsymbol{\mu}_1 = 0$, remaining resources $\boldsymbol{\beta}_1 = \bm{\gamma}$, and momentum gradient $\mathbf{g}_0 = 0$.
	    \FOR{$t=1,\cdots,T$}
    	    \STATE User $u_{nT+t}$ arrives
    	    \STATE 
    	    $
            \mathbf{m}_p= \left \{
            \begin{array}{ll}
                0,                    & \boldsymbol{\beta}_{tp} > 0\\
                \infty,                    & \textrm{otherwise}
            \end{array}
            \right.
            $
            
    	    \STATE  $// ~~ \texttt{Make the recommendation:}$
    	    \STATE 
    	    $
    	        \mathbf{x}_t = \argmax_{\mathbf{x}_t\in\mathcal{X}}\left[g(\mathbf{x}_t)/T-\left(\mathbf{A}(\boldsymbol{\mu}_t+\mathbf{m})\right)^{\top}\mathbf{x}_t\right]
    	    $
    	    \STATE $\boldsymbol{\beta}_{t+1} = \boldsymbol{\beta}_{t} - \mathbf{A}^{\top}\mathbf{x}_t$
    	    
    	    \STATE $\mathbf{e}_t = \argmax_{\mathbf{e_t}\leq\boldsymbol{\beta}_t}r^*(-\mathbf{\boldsymbol{\mu}_t})$
            \STATE
           $
                \widetilde{\mathbf{g}}_t = -\mathbf{A}^{\top}\mathbf{x}_t + \mathbf{e}_t
            $
            \STATE
            $
                \mathbf{g}_t =\alpha \widetilde{\mathbf{g}}_t + (1-\alpha)\mathbf{g}_{t-1}
            $
            \STATE
            $
                \boldsymbol{\mu}_{t+1} = \argmin_{\boldsymbol{\mu}\in\mathcal{D}}\left[\langle \mathbf{g}_t,\boldsymbol{\mu} \rangle + \eta \| \boldsymbol{\mu}-\boldsymbol{\mu}_t \|_{\boldsymbol{\gamma}^2}^2\right]
            $
    	\ENDFOR
	\ENDFOR
	\
	\end{algorithmic}
\end{algorithm}


\section{Experiment}\label{sec:exp}
We conducted experiments to show the effectiveness of the proposed P-MMF for provider-fair recommendations. 
The source code and experiments have been shared at github~\footnote{\url{https://github.com/XuChen0427/P-MMF}. For the MindSpore~\cite{mindspore} version, please see~\url{https://gitee.com/mindspore/models/tree/master/research/recommend/pmmf}}.

\subsection{Experimental settings}
\subsubsection{Datasets}
The experiments were conducted on four large-scale, publicly available recommendation datasets, including:
 
 \textbf{Yelp}\footnote{\url{https://www.yelp.com/dataset}}: a large-scale businesses recommendation dataset. We only utilized the clicked data, which is simulated as the 4-5 star rating samples. The cities of the items are considered as providers. It has 154543 samples, which contains 17034 users, 11821 items and 23 providers.

 \textbf{Amazon-Beauty/Amazon-Baby}: Two subsets (beauty and digital music domains) of Amazon Product dataset\footnote{\url{http://jmcauley.ucsd.edu/data/amazon/}}. We only utilized the clicked data, which is simulated as the 4-5 star rating samples. Also, the brands are considered as providers. They has 49217/59836 samples, which contains 9625/11680 users, 2756/2687 items and 1024/112 providers.
 
 \textbf{Steam}\footnote{\url{http://cseweb.ucsd.edu/~wckang/Steam_games.json.gz}}~\cite{SASRec}: We used the data for gamed played for more than 10 hours in our experiments. The publishers of games are considered as providers. It has 29530 samples, which contains 5902 users, 591 items and 81 providers.

As a pre-processing step, the users, items, and providers who interacted with less than 5 items/users were removed from all dataset to avoid the extremely sparse cases. We also removed the providers associated with less than 5 items.


\subsubsection{Evaluation}
We sorted all the interactions according to the time and used the first 80\% of the interactions as the training data to train the base model (i.e., BPR~\cite{BPR}). The remaining 20\% of interactions were used as the test data for evaluation. Based on the trained base model, we can obtain a preference score $s_{u, i}$ for each user-item pair $(u, i)$.
The chronological interactions in the test data were split into interaction sequences where the horizon length was set to $T$. We calculated the metrics separately for each sequence, and the averaged results are reported as the final performances.

As for the evaluation metrics, the performances of the models were evaluated from three aspects: user-side preference, provider-side fairness, and the trade-off between them. As for the user-side preference, following the practices in~\cite{wu2021tfrom}, we utilized the NDCG@K, which is defined as the ratio between the sum of position-based user-item scores~\cite{wu2021tfrom} in the original ranking list $\mathbf{L}_K(u_t)$ and that in the re-ranked list $\mathbf{L}^F_K(u_t)$:
$$
\text{NDCG@K} =\frac{1}{T} \sum_{t=1}^T\frac{\sum_{i\in\mathbf{L}_K(u_t)}s_{u_t,i}/\log(\textrm{rank}_i+1)}{\sum_{i\in\mathbf{L}_K^F(u_t)}s_{u_t,i}/\log(\textrm{rank}_i^F+1)},
$$
where rank$_i$ and rank$_i^F$ are the ranking positions of the item $i$ in $\mathbf{L}_K(u_t)$ and $\mathbf{L}_K^F(u_t)$, respectively.

As for the provider fairness, we directly utilized the definition of MMF in Section~\ref{sec:formulation} as the metric:
\[
    \textrm{MMF}@K = \min_{p\in\mathcal{P}}\left\{\sum_{t=1}^T\sum_{i\in\mathbf{L}_K^F(u_t)}\mathbb{I}(i\in\mathcal{I}_p)/\boldsymbol{\gamma}_p\right\},
\]
where $\mathbb{I}(\cdot)$ is the indicator function. 

As for the trade-off performance, we used the online objective traded-off with Equation~\eqref{eq:W_eq}:
\[
    W_{\lambda}@K = \frac{1}{T}\sum_{t=1}^T\left(\sum_{i\in\mathbf{L}_K^F(u_t)}s_{u_t,i}\right) + \lambda\cdot \textrm{MMF}@K,
\]
where $\lambda\geq 0$ is the trade-off coefficient.

\subsubsection{Baselines}
The following representative provider fair re-ranking models were chosen as the baselines:
\textbf{FairRec}~\cite{fairrec} and \textbf{FairRec+} \cite{fairrecplus} aimed to guarantee at least Maximin Share (MMS) of the provider exposures.
\textbf{CPFair}~\cite{cpfair} formulated the trade-off problem as a knapsack problem and proposed a greedy solution.

We also chose the following MMF models:
\textbf{Welf}~\cite{nips21Welf} use the Frank-Wolfe algorithm to maximize the Welfare functions of worst-off items. However, it is developed under off-line settings; \textbf{RAOP}~\cite{balseiro2021regularized} is a state-of-the-art online resource allocation method. We applied it to the recommendation by regarding the items as the resources and users as the demanders. 

We also compared the proposed P-MMF with two heuristic MMF baselines:
\textbf{$K$-neighbor}: at each time step $t$, only the items associated to the top-$K$ providers with the least cumulative exposure are recommended;  \textbf{min-regularizer}: at each time step $t$, we add a regularizer that measures the exposure gaps between the target provider and the worst-providers. Appendix~\ref{app:min-regularizer} shows the algorithm details.


\subsubsection{Implementation details}
As for the hyper-parameters in all models, the learning rate was tuned among $[1e-2,1e-4]/T^{1/2}$ and the momentum coefficient $\alpha$ was tuned among $[0.2,0.6]$. For the maximum resources (i.e., the weights) $\bm{\gamma}$, following the practices in~\cite{wu2021tfrom}, we set $\bm{\gamma}$ based on the number of items provided by the providers:
$
    \bm{\gamma}_p = KT\eta{|\mathcal{I}_p|}\big/{|\mathcal{I}|},
$
where $\eta$ is the factor controlling the richness of resources. In all the experiments, we set $\eta=1+1/|\mathcal{P}|$. We implemented P-MMF with both CPU and GPU versions based on cvxpy~\cite{cvxpy} and its PyTorch version~\cite{agrawal2019differentiable}, respectively. 



\subsection{Simulation on a small dataset}\label{sec:exp_simulation}
Note that Problem~\eqref{eq:solve_OPT} can use MMF or PF as its regularizer, though directly solving it on large-scale datasets is difficult. 
To verify the correctness of the formulation and to investigate different impacts of MMF-based and PF-based regularizers on provider fair recommendation, we first conducted a numerical simulation using 5\% of the Yelp data, which consists of 844 users, 813 items, 10 providers, and with the length of the horizon $T=256$ . 

Specifically, we solved the Problem~\eqref{eq:solve_OPT} with PF based or MMF based regularizers (i.e., setting their provider weights as even distribution $\bm{\gamma}_p = KT, \forall p\in \mathcal{P}$), using the solver cvxpy~\cite{cvxpy}.
At each time $t$, after receiving the recommendation decision variable $\mathbf{x}_t$, we calculated the overall provider exposure $\mathbf{e}$ on the dataset. Then the Lorenz curves~\cite{Lorenzcurve} of provider exposures were drawn and shown in Figure~\ref{fig:simulation}. The Lorenz curve is often used to represent exposure distribution. Here it shows the proportion of overall exposure percentage assumed by the bottom $x$\% providers. In other words, for the bottom $x$\% providers, what percentage ($y$\%) of the total exposures they have. Note that the diagonal line to the upper right is known as the absolute fair line.

From the curves shown in Figure~\ref{fig:simulation}, we can observe that when the $\lambda \rightarrow \infty$ (i.e., only consider provider's fairness), both PF and MMF can achieve the expected proportion $\bm{\gamma}$ (i.e., even exposures here). However, after considering the user preference, the MMF-based regularizer tends to consider the worst providers' exposure first, while PF does not. For example, when the trade-off coefficient $\lambda$ changes among $[0.01,0.05,0.1]$, 60\% of the least exposure providers will increase $[18.81\%,18.96\%, 12.58\%]$ exposures by MMF. If using PF-based regularizer, then the increased exposure ratios become $[10.1\%,16.27\%, 9.58\%]$. The results verified that formulating recommendations as a resource allocation problem regularized by MMF leads to better provider-fair recommendations than PF. 

\begin{figure}[t]  
    
    \centering    
    \subfigure[PF]
    {
        \includegraphics[width=0.4\linewidth]{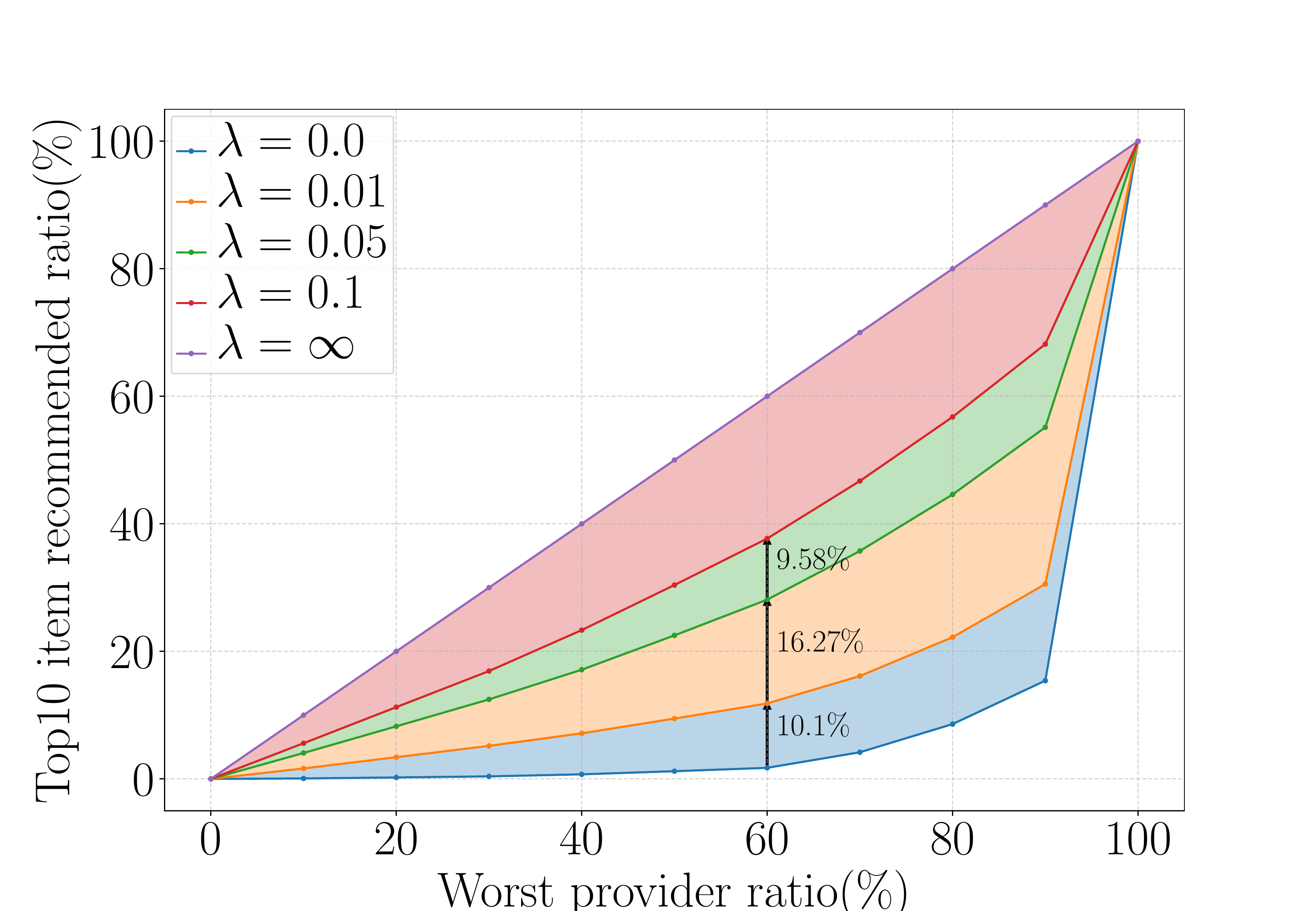}
    }
    \subfigure[MMF]
    {
        \includegraphics[width=0.4\linewidth]{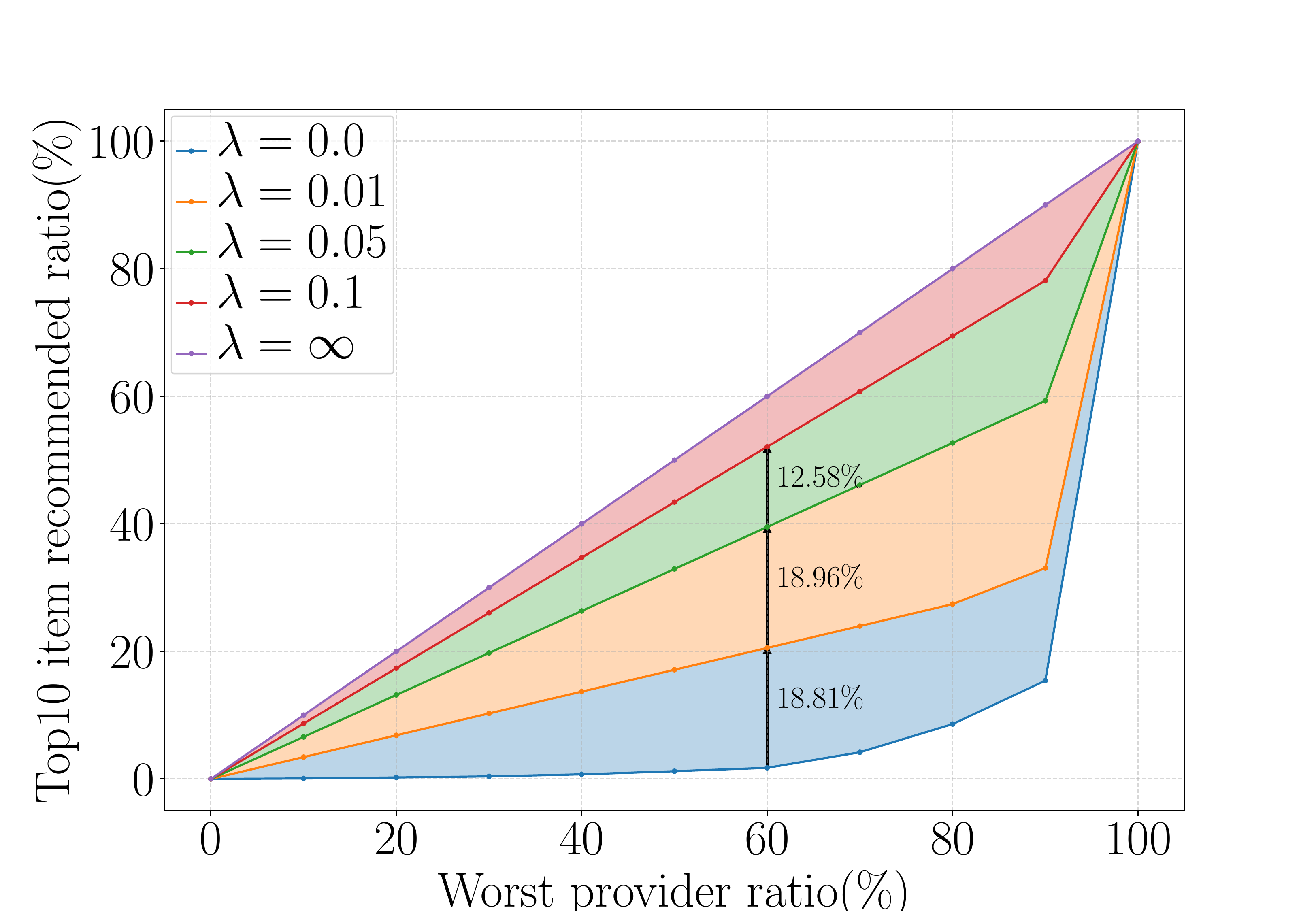}
    }
  
    \caption{The Lorenz curves of provider exposure with PF regularizer (a) and MMF regularizer (b), respectively. The experiments were conducted based on 5\% of the Yelp dataset. }
    \label{fig:simulation}
   \vspace{-0.3cm}
\end{figure}


\subsection{Experimental results on full datasets}
In this section, we conducted the experiments with the online algorithm developed in Section~\ref{sec:P-MMF}. In all of the experiments, BPR~\cite{BPR} was chosen as the base ranking model for generating preference scores. We set the length of the horizon $T=256$.

Table~\ref{tab:EXP:main} reports the experimental results of P-MMF and the baselines on all datasets in terms of the metric $W_{1}@K$. Underlined numbers mean the best-performed baseline. To make fair comparisons, all the baselines were tuned and used $W_{1}@K$ as the evaluation metric. Note that similar experiment phenomena has also been observed on other $\lambda$ values.

From the reported results, we found that P-MMF outperformed all of the PF-based baselines in terms of $W_{1}@K$ ($K=5, 10, 20$), verified that P-MMF can give supports to the poor-conditioned providers. We also observed that P-MMF outperformed all the MMF-based baselines, indicating P-MMF's effectiveness in enhancing provider fairness while keeping high user preference.


\begin{table*}[ht]
\setlength{\tabcolsep}{4.5pt}
        \small
        \caption{Performance comparisons between ours and the baselines on Yelp, Beauty, Baby, and Steam. We choose trade-off co-efficient $\lambda=1$ on three different ranking sizes $K$ to investigate the effectiveness of ours.}
    \label{tab:EXP:main}
    \centering
    \centering
    \begin{tabular}{|l|l|rrr|rrrr|rr|}
        \hline
        \multicolumn{2}{|c|}{} & \multicolumn{3}{c|}{PF-based baselines} & \multicolumn{4}{c|}{MMF-based baselines} & \multicolumn{2}{c|}{Our approach} \\
        \hline
         \textbf{Dataset} &{Metric} & FairRec & FairRec+  & CPFair & k-neighbor & Welf & min-regularizer & ROAP & \textbf{P-MMF(ours)} & Improv.\\
        \hline
\multirow{3}{*}{Yelp}  & {$W_{1}@$5} & 2.659 & 2.955  & 3.366 & 2.929 & 2.986 & 3.296 & \underline{3.395} &  \textbf{3.455} & 1.8\% \\
& {$W_{1}@$10} & 5.046 & 5.335  & 6.124 & 5.345 & 5.639 & 6.125 & \underline{6.126} &  \textbf{6.247} & 2.0\% \\
& {$W_{1}@$20}& 10.256 & 10.860  & 11.370 & 10.829 & \underline{11.648} & 11.600 & 11.640 &  \textbf{11.797} & 1.3\% \\

\hline
\multirow{3}{*}{Amazon-Beauty} & {$W_{1}@$5} & 2.617 & 3.783  & \underline{4.959} & 3.958 & 4.820 & 4.953 & 4.863 &  \textbf{5.034} & 1.5\% \\
& {$W_{1}@$10} & 5.197 & 7.244  & 9.195 & 7.312 & 9.166 & \underline{9.309} & 9.147 &  \textbf{9.434} & 1.3\% \\
& {$W_{1}@$20}& 10.312 & 13.492  & 17.686 & 13.474 & 17.650 & \underline{17.880} & 17.637 &  \textbf{17.983} & 0.6\% \\

\hline
\multirow{3}{*}{Amazon-Baby} & {$W_{1}@$5} & 2.696 & 3.377  & 4.129 & 3.438 & 4.344 & \underline{4.556} & 4.303 &  \textbf{4.583} & 0.6\% \\
& {$W_{1}@$10} & 5.388 & 6.499  & 8.080 & 6.384 & 8.187 & \underline{8.345} & 8.137 &  \textbf{8.440} & 1.1\% \\
& {$W_{1}@$20}& 10.591 & 12.193  & 15.527 & 11.871 & 15.480 & \underline{15.713} & 15.544 &  \textbf{15.873} & 1.1\% \\

\hline
\multirow{3}{*}{Steam} & {$W_{1}@$5} & 2.953 & 3.173  & 3.994 & 2.992 & 3.947 & \underline{4.005} & 3.915 &  \textbf{4.443} & 10.9\% \\
& {$W_{1}@$10} & 5.486 & 5.716  & 7.736 & 5.479 & 7.678 & \underline{8.033} & 7.893 &  \textbf{8.148} & 1.4\% \\
& {$W_{1}@$20} & 9.915 & 12.699  & 14.864 & 9.724 & 14.733 & 15.039 & \underline{15.081} &  \textbf{15.235} & 1.0\% \\
\hline

\hline
    \end{tabular}
\end{table*}

Figure~\ref{fig:Pareto_bound} shows the Pareto frontiers~\cite{lotov2008visualizing} of NDCG@K and MMF@K on four datasets with different ranking size $K$. The Pareto frontiers were drawn by tuning different parameters of the models and choosing the (NDCG@K, MMF@K) points with the best performances. In the experiment, we selected the baselines of CPFair, min-regularizer, Welf, and ROAP, which achieved relatively good performances among all baselines.

From the Pareto frontiers, we can see that the proposed P-MMF Pareto dominated the baselines (i.e., the P-MMF curves are at the upper right corner). Pareto dominance means that under the same NDCG@K level, P-MMF achieved better MMF@K; Under the same MMF@K level, P-MMF achieved better NDCG@K. The results demonstrate that P-MMF splendidly improves the utilities of poor-conditioned providers without sacrificing users' utilities too much. 

\begin{figure*}
        \centering    
    \subfigure[Amazon Beauty $K=5$]
    {
        \includegraphics[width=0.2\linewidth]{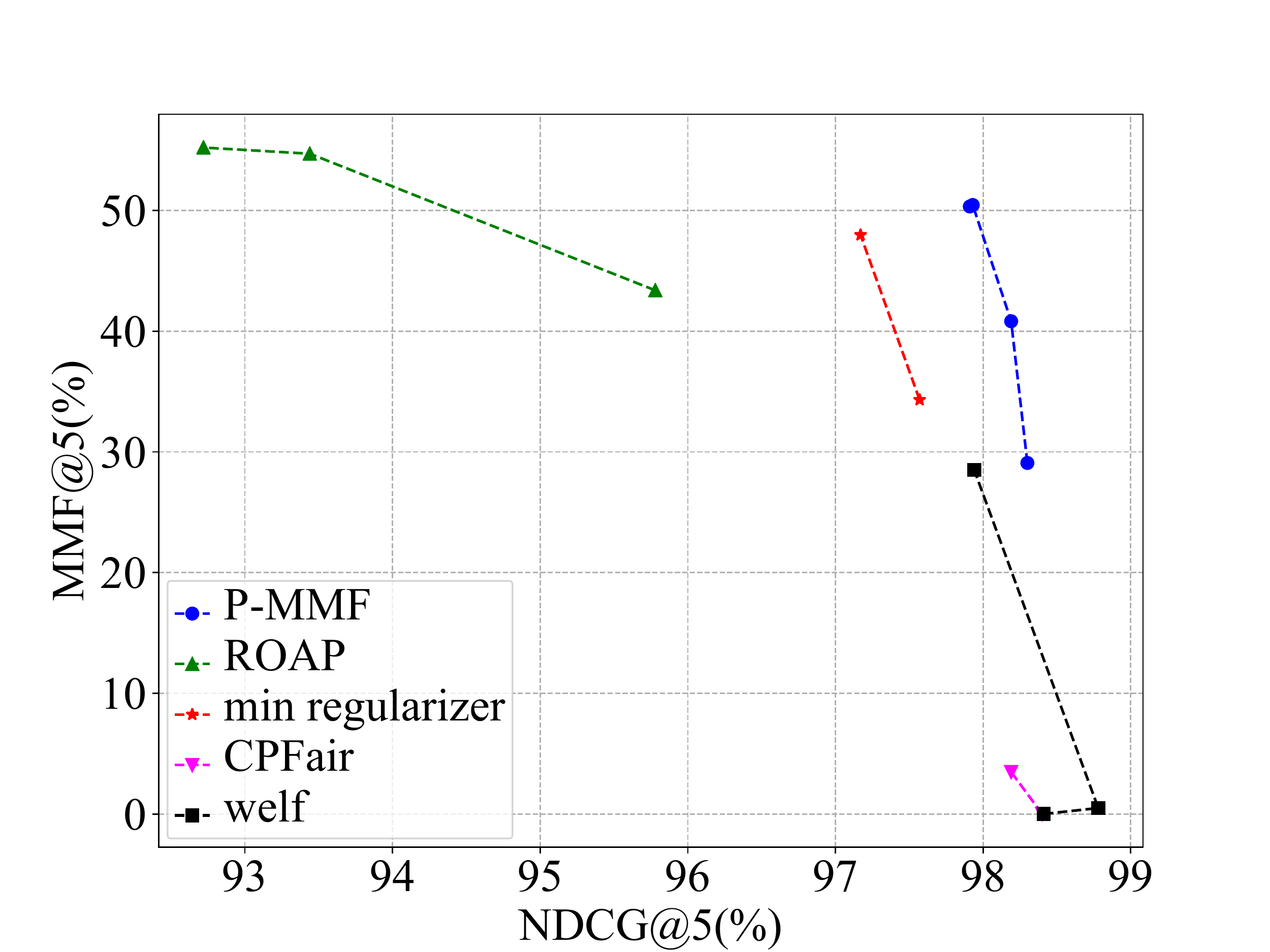}
    }
     \subfigure[Amazon Baby $K=5$]
    {
        \includegraphics[width=0.2\linewidth]{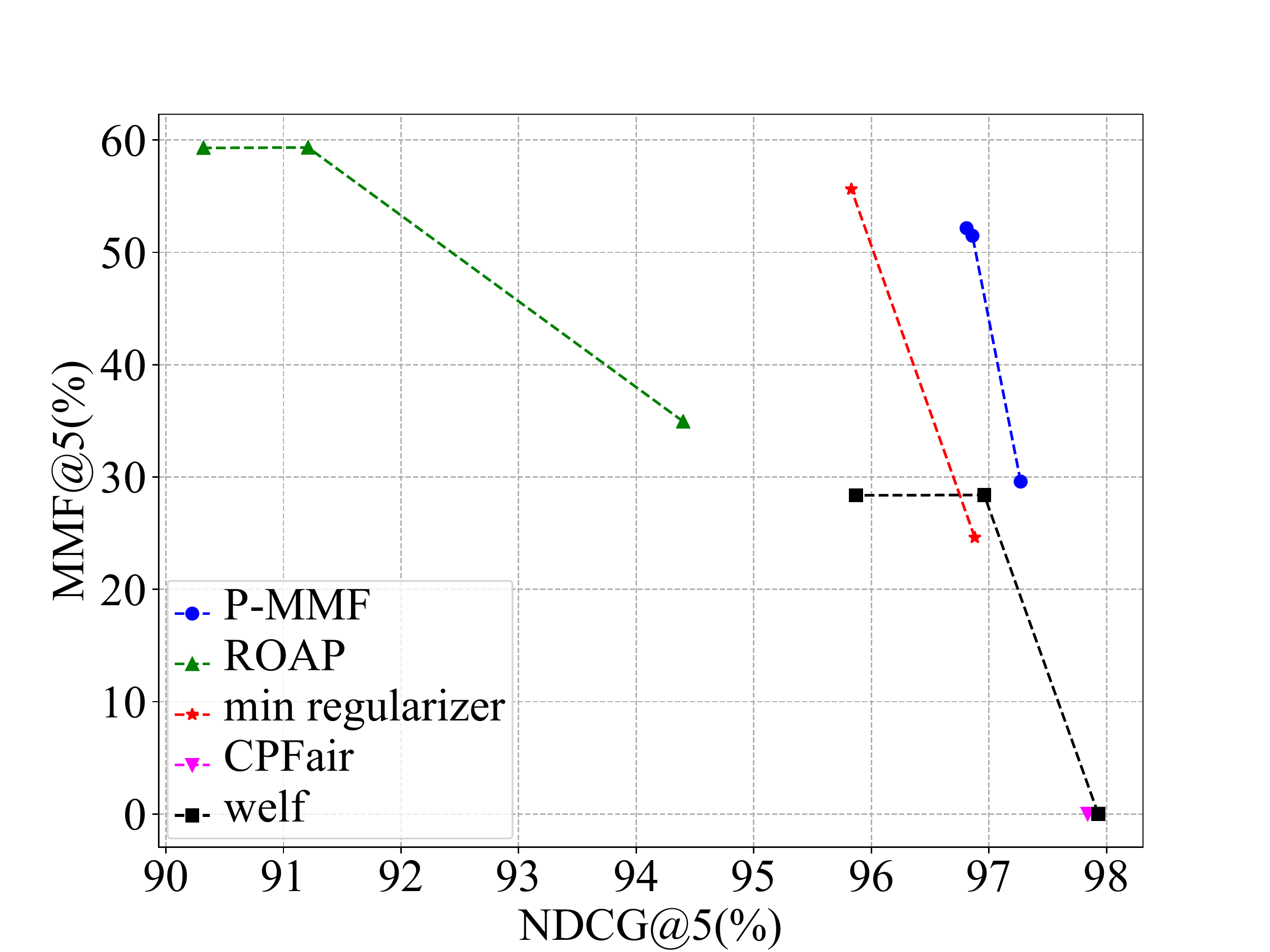}
    }
     \subfigure[Yelp $K=5$]
    {
        \includegraphics[width=0.2\linewidth]{image/Amazon_Beauty_bound_K5T256_figure.pdf}
    }
     \subfigure[Steam $K=5$]
    {
        \includegraphics[width=0.2\linewidth]{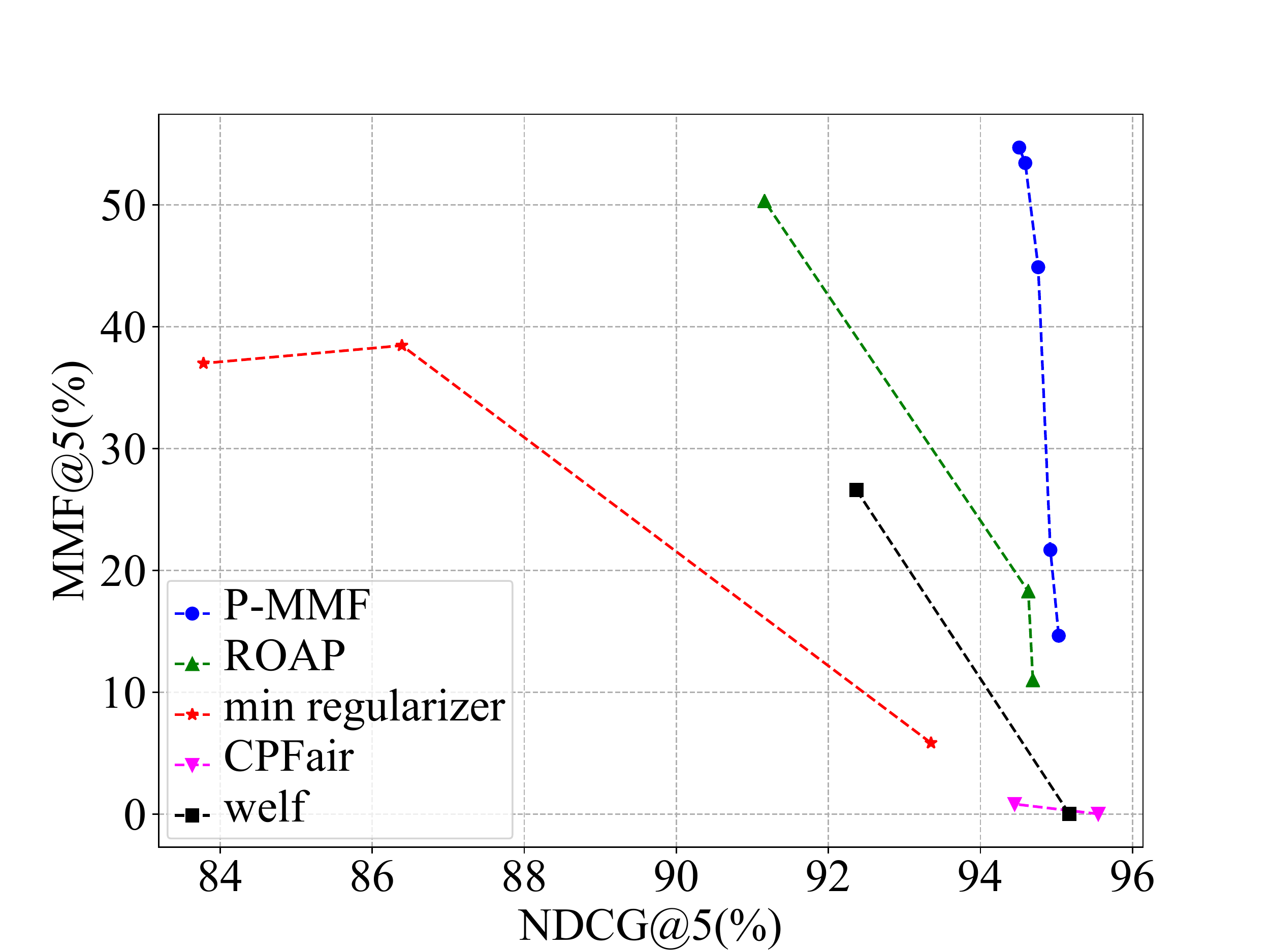}
    }
    
    \subfigure[Amazon Beauty $K=10$]
    {
        \includegraphics[width=0.2\linewidth]{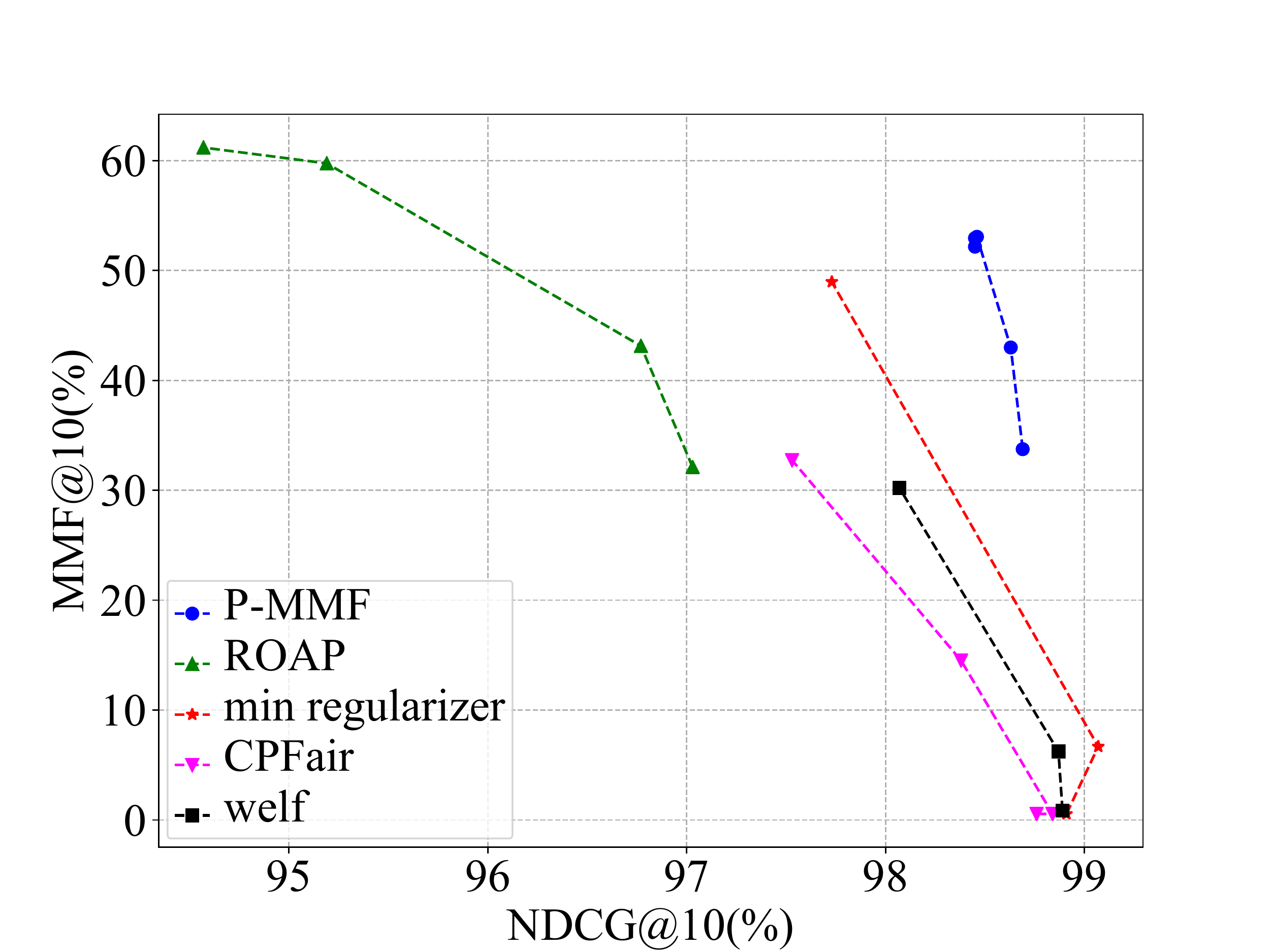}
    }
     \subfigure[Amazon Baby $K=10$]
    {
        \includegraphics[width=0.2\linewidth]{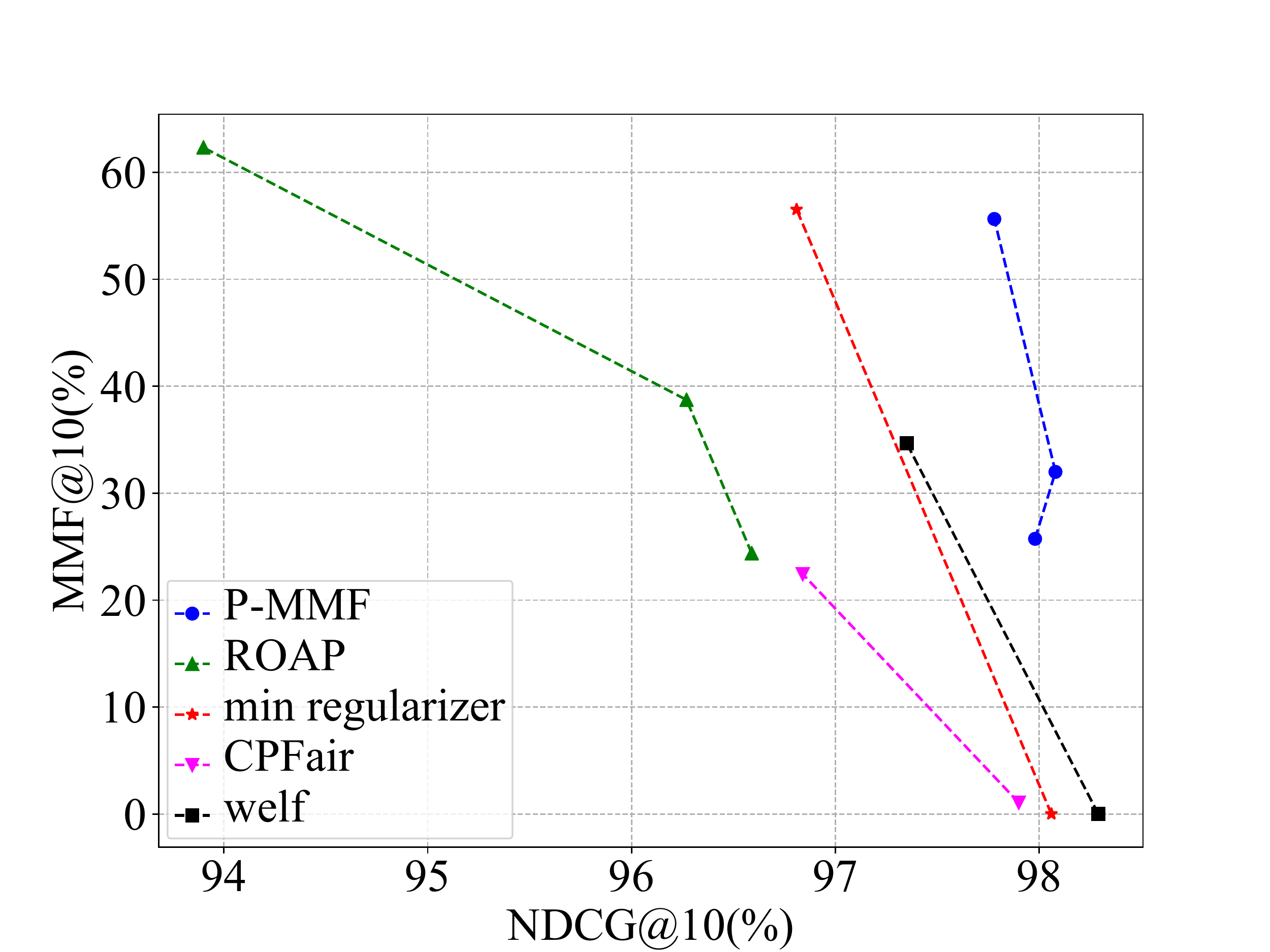}
    }
     \subfigure[Yelp $K=10$]
    {
        \includegraphics[width=0.2\linewidth]{image/Amazon_Beauty_bound_K10T256_figure.pdf}
    }
     \subfigure[Steam $K=10$]
    {
        \includegraphics[width=0.2\linewidth]{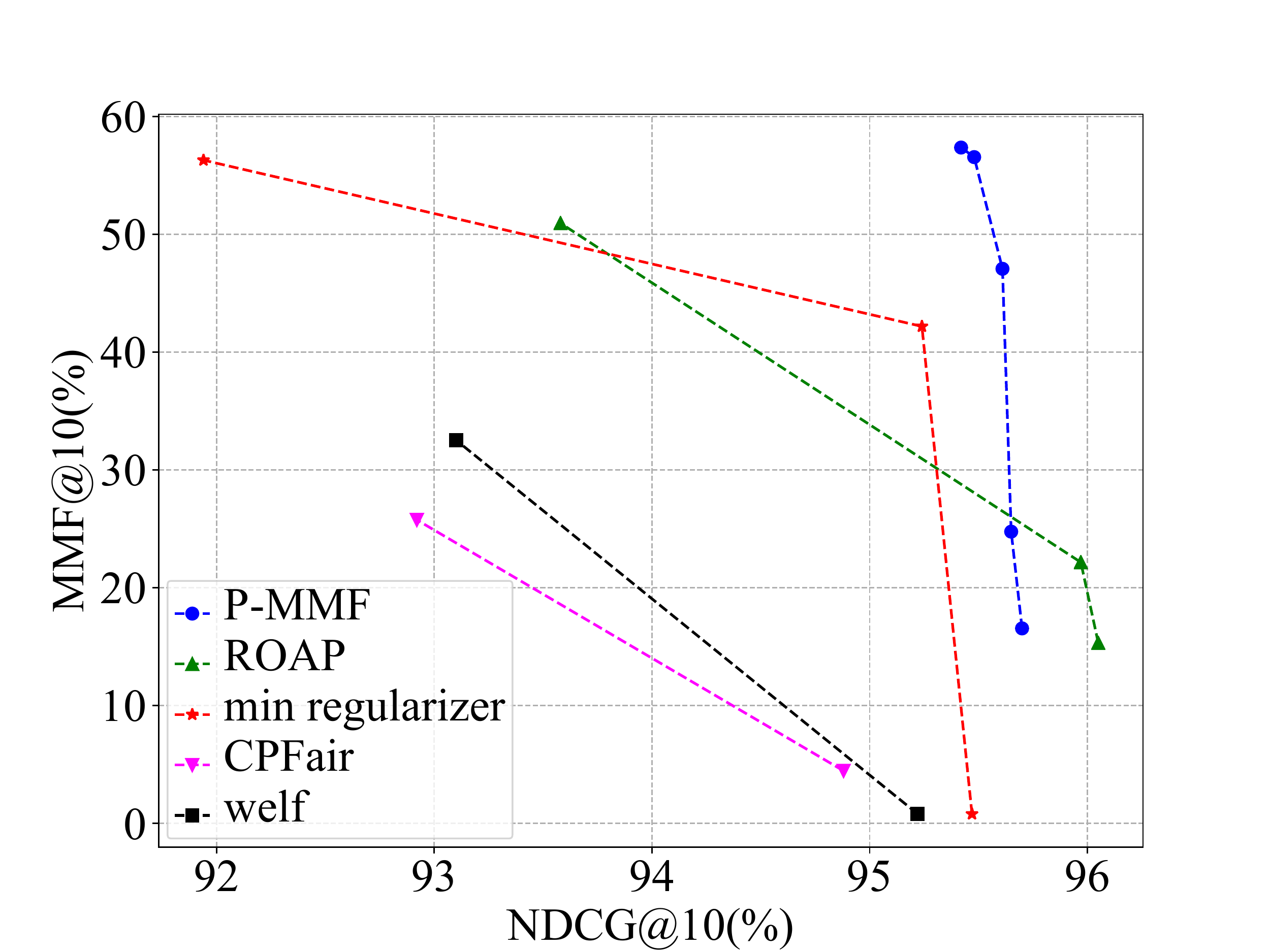}
    }
    
     \subfigure[Amazon Beauty $K=20$]
    {
        \includegraphics[width=0.2\linewidth]{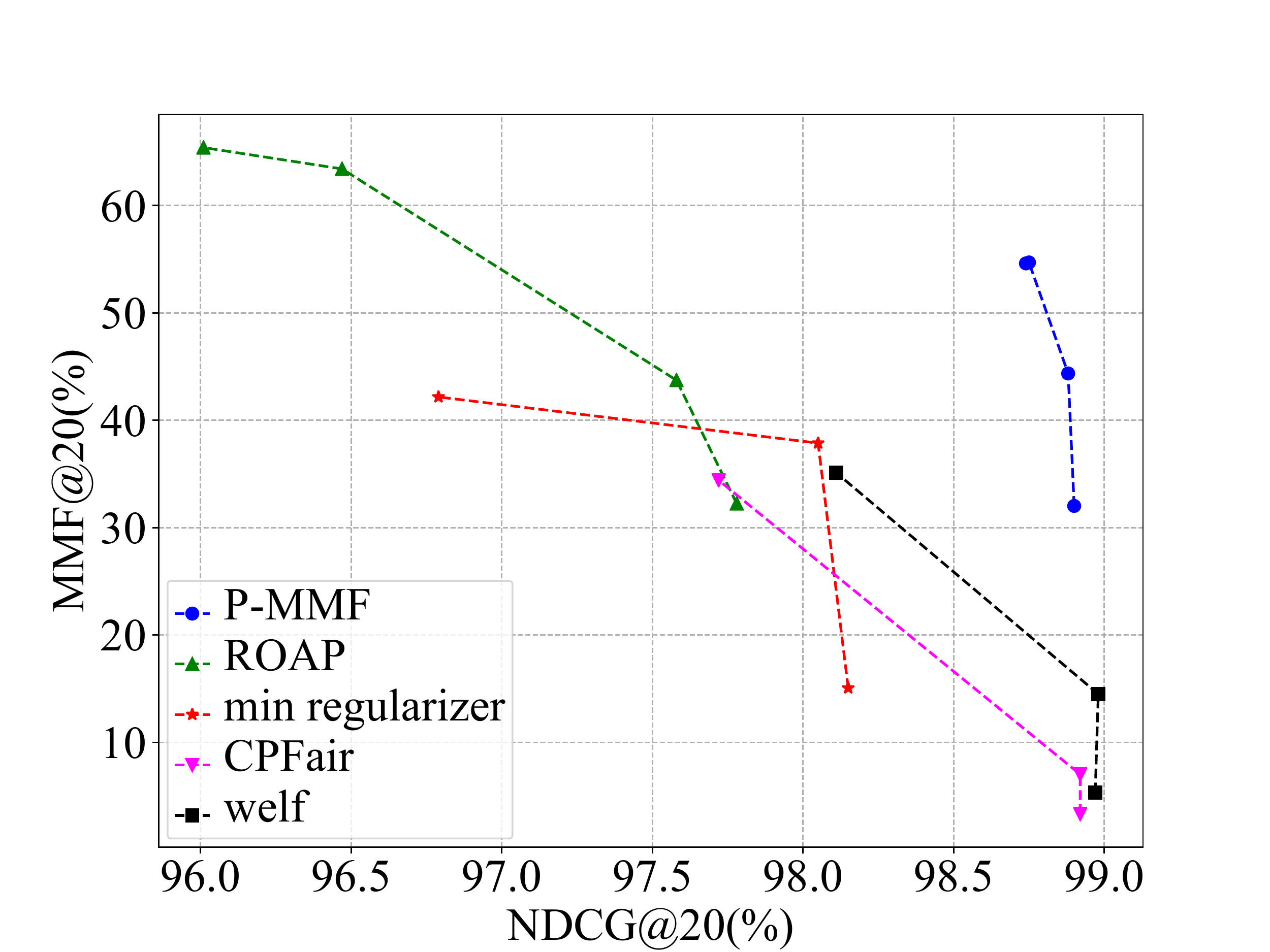}
    }
     \subfigure[Amazon Baby $K=20$]
    {
        \includegraphics[width=0.2\linewidth]{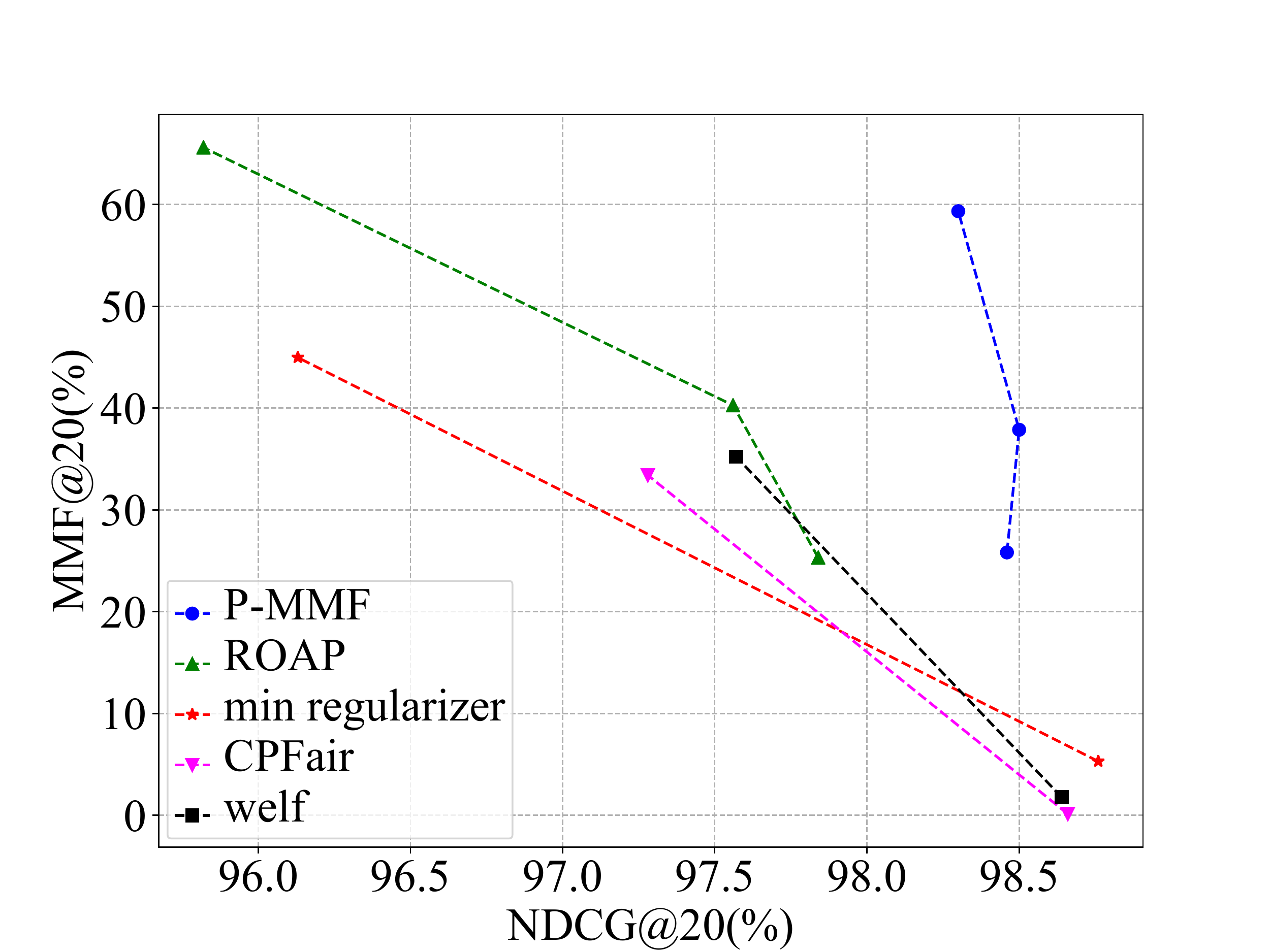}
    }
     \subfigure[Yelp $K=20$]
    {
        \includegraphics[width=0.2\linewidth]{image/Amazon_Beauty_bound_K20T256_figure.pdf}
    }
     \subfigure[Steam $K=20$]
    {
        \includegraphics[width=0.2\linewidth]{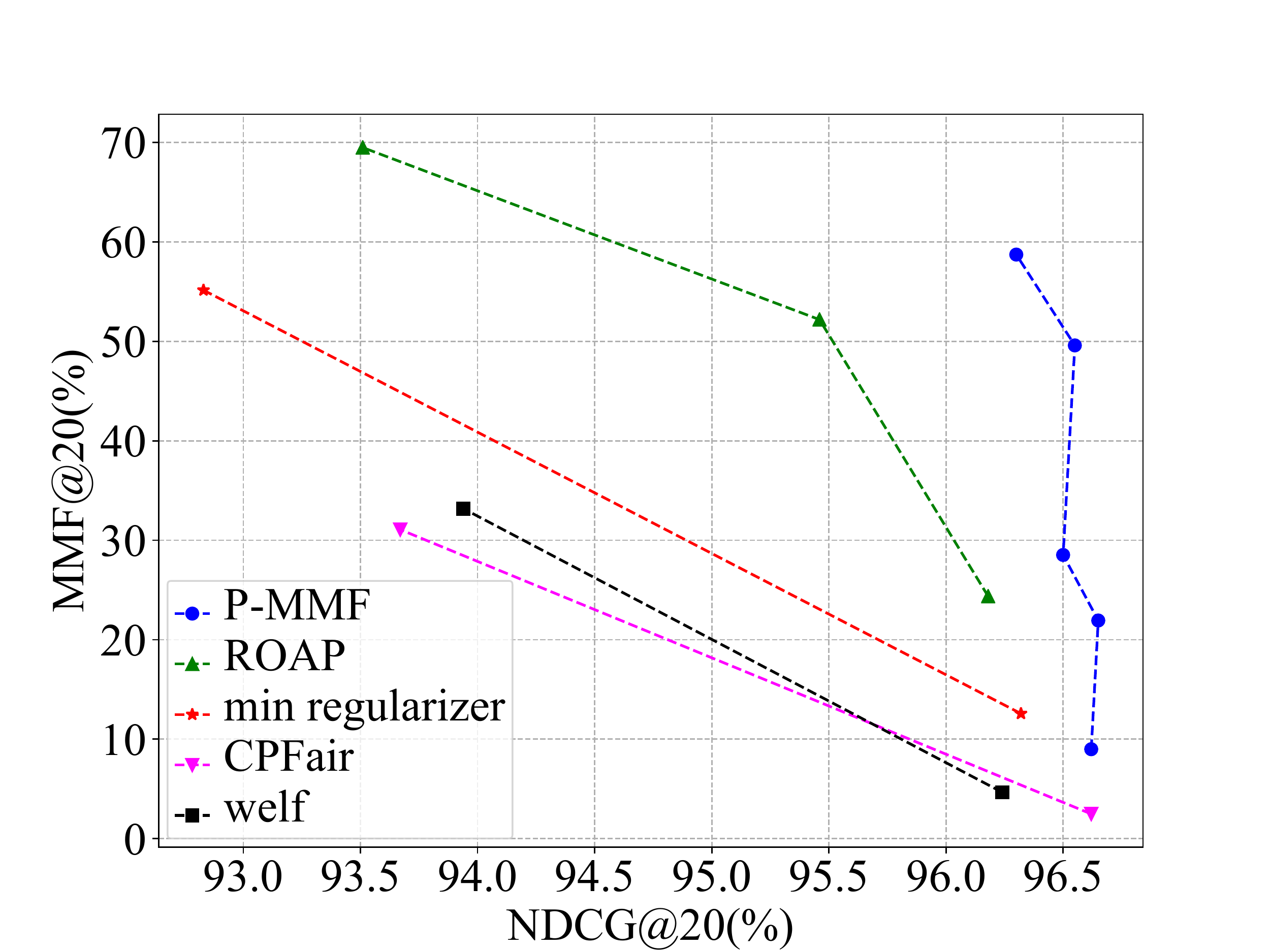}
    }
    \caption{Pareto frontier of four different datasets with different top-$K$ ranking. }
    \label{fig:Pareto_bound}
\end{figure*}

\begin{table*}[ht]
\setlength{\tabcolsep}{4.5pt}
        \small
        \caption{Performance comparisons between ours and the baselines on Yelp with different base recommender models. We choose trade-off co-efficient $\lambda=1$ on three different ranking sizes $K$ to investigate the effectiveness of ours.}
    \label{tab:EXP:main2}
    \centering
    \centering
    \begin{tabular}{|r|l|rrr|rrrr|rr|}
        \hline
        \multicolumn{2}{|c|}{} & \multicolumn{3}{c|}{PF-based baselines} & \multicolumn{4}{c|}{MMF-based baselines} & \multicolumn{2}{c|} {Our approach}\\
        \hline
         \textbf{Base Model} &{Metric} & FairRec & FairRec+  & CPFair & k-neighbor & Welf & min-regularizer & ROAP & \textbf{P-MMF(ours)} & Improv.\\
        \hline

\multirow{3}{*}{DMF~\cite{DMF}} & {$W_{1}@$5} & 3.1982 & 4.087  & 5.070 & 4.114 & 4.981 & \underline{5.198} & 5.190 &  \textbf{5.272} & 1.4\% \\
& {$W_{1}@$10} & 6.364  & 7.829  & 9.929 & 7.957 & 9.867 & \underline{9.968} & 9.878 & \textbf{10.003} & 0.3\% \\
& {$W_{1}@$20}& 12.643 & 15.148  & 19.372 & 15.208 & 19.292 & \underline{19.410} & 19.324 &  \textbf{19.448} & 0.2\% \\

\hline
\multirow{3}{*}{LINE~\cite{LINE}} & {$W_{1}@$5} & 2.323 & 4.074  & 5.237 & 4.269 & 5.153 & \underline{5.239} & 5.159 &  \textbf{5.407} & 3.2\% \\
& {$W_{1}@$10} & 4.614 & 7.866 & 10.087 & 8.099 & 10.064 & \underline{10.124} & 10.042 &  \textbf{10.280} & 1.5\% \\
& {$W_{1}@$20}& 9.164 & 15.041 & 19.814 & 15.829 & \underline{19.826} & 19.450 & 19.723 &  \textbf{19.973} & 0.7\% \\

\hline
\multirow{3}{*}{LightGCN~\cite{ligntGCN}} & {$W_{1}@$5} & 2.729 & 3.714 & 4.531 & 3.963 & 4.633 & \underline{4.782} & 4.718 &  \textbf{4.837} & 1.2\% \\
& {$W_{1}@$10} & 5.468 & 7.2368  & 8.982 & 7.584 & 8.975 & 9.016 & \underline{9.144} & \textbf{9.283} & 1.5\% \\
& {$W_{1}@$20} & 10.927 & 13.995 & 17.799 & 14.856 & 17.774 & 17.812 & \underline{17.871} & \textbf{18.035} & 0.9\% \\
\hline
\end{tabular}
\end{table*}

\subsection{Experiment analysis}
We also conducted experiments to analyze P-MMF on Yelp. 


\subsubsection{Ablation study on different base models}\label{sec:ab4BM}
P-MMF and other baselines are re-ranking models which re-rank the results outputted by a base recommender model. To verify the effectiveness of P-MMF with different base ranking models, we choose three other base models to generate user-item preference scores $s_{u,i}$, including \textbf{DMF}~\cite{DMF} which optimizes the matrix factorization with the deep neuarl networks; \textbf{LINE}~\cite{LINE} is a matrix factorization model based on graph embeddings; and \textbf{LightGCN}~\cite{ligntGCN} which builts a user-item interaction graph and adapts Graph Convolutional Network~\cite{zhang2019graph} to conduct recommendation. All the experiments were also conducted on the full Yelp dataset with $T=256$.

Table~\ref{tab:EXP:main2} shows the experimental results of P-MMF with different base models. We observed that P-MMF consistently outperformed the PF-based and MMF-based baselines. The results also verified that P-MMF is more effective than the baselines in re-ranking the results outputted by different base models.


\subsubsection{Regret bound experiments}\label{sec:reg}
To further show the effectiveness of the P-MMF, we directly compute the regret Regret$(h)=W_{OPT} - W$, where $W_{OPT}$ is the oracle performance defined in Equation~\eqref{eq:OPT}, and $W$ is the online performance defined in Equation~\eqref{eq:W_eq}. Due to the huge computational cost of obtaining $W_{OPT}$ on large-scale datasets, we conducted the experiments on the 5\% of the Yelp data created in Section~\ref{sec:exp_simulation}. We compared P-MMF with the state-of-the-art provider-fair online MMF baselines: min-regularizer and ROAP. The experiments were conducted on ranking size $K=10$. Note that in the experiment, we fixed the user arriving size $N$, and the regret is computed through the summation over the $N/T$ samples where $T$ is the length of the horizon. According the Theorem~\ref{theo:regret}, the summation Regret$(h)$ is comparable with $O(T^{1/2}N/T) = O(N/T^{1/2})$.

Figure~\ref{fig:regret} shows the summation Regret$(h)$ curves w.r.t. $T$. Figure~(a) and (b) show the curves when the trade-off co-efficient $\lambda$ was set as $1$ and $0.1$, respectively. 
From Figure~\ref{fig:regret}, we can see that P-MMF has lower regret bound than other online models, especially when $\lambda$ is large. Moreover, we can see that the regret bound of P-MMF is decreasing with the increase of $T$, which verified the theoretical analysis results.
Similar results have also been observed for other $\lambda$ values.

\begin{figure}[t]  
    \centering    
    \subfigure[$\lambda=1$]
    {
        \includegraphics[width=0.4\linewidth]{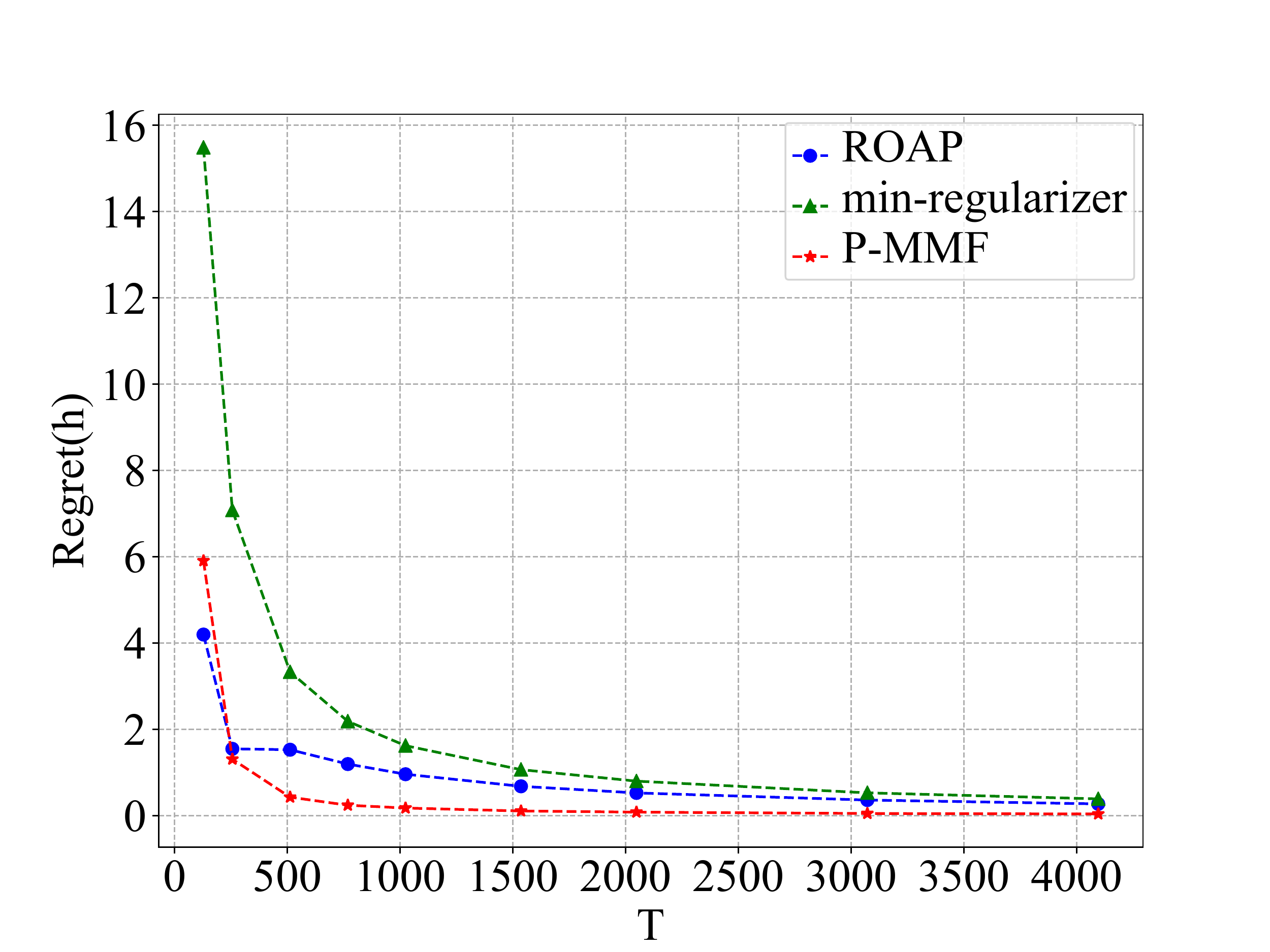}
    }
    \subfigure[$\lambda=1e-1$]
    {
        \includegraphics[width=0.4\linewidth]{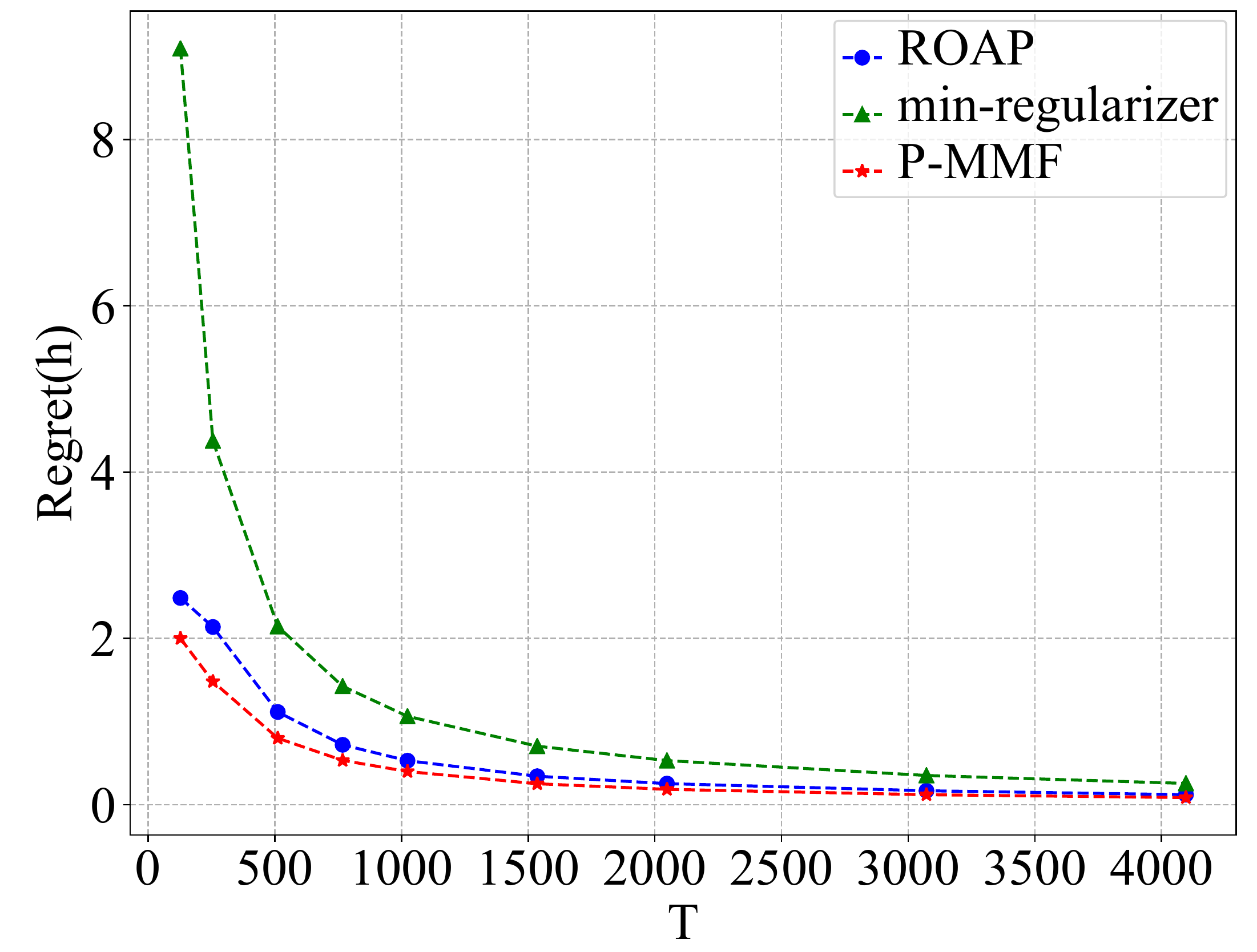}
    }
    \caption{Regret versus time-separate length $T$.}
    \label{fig:regret}  
\end{figure}

\subsubsection{Online inference time}
We experimented with investigating the online inference time of P-MMF and the most practical RS model DMF~\cite{DMF}. 
In real recommender systems, the number of providers is relatively small and steady. However, the number of users and items is usually huge and grows rapidly. Therefore, we tested the inference time under CPU and GPU implementations of P-MMF and DMF w.r.t. the different number of items while keeping the number of users and providers unchanged. The GPU implementation is based on PyTorch~\cite{pytorch}.

Figure~\ref{fig:inference_time} reports the curves of inference time (ms) per user access w.r.t. item size. We can see that P-MMF with CPU and GPU versions need only about 20-40ms and 17-18ms for online inference, respectively. Moreover, we can see that the inference time of P-MMF did not increase much with the increasing number of items. For example, by increasing the item size from 0 to 200000, the P-MMF CPU version only needs a little bit more time (19ms) for online inference. The inference time for the GPU version almost kept unchanged. As for comparisons, DMF's inference time increased rapidly: both CPU and GPU versions need much more time (about 65ms). The phenomenon can be easily explained with the dual problem analysis in Theorem~\ref{theo:dual}. We see the parameter size of P-MMF is provider size, which is far less than the item size $|\mathcal{P}|\ll |\mathcal{I}|$. Therefore, the online inference time is not sensitive to item numbers. We conclude that P-MMF can be adapted to the real online recommendation scenarios efficiently because of its low and robust online computational cost, even when the number of items grows rapidly.

\begin{figure}[t]  
    \centering    
    \subfigure[CPU version]
    {
        \includegraphics[width=0.4\linewidth]{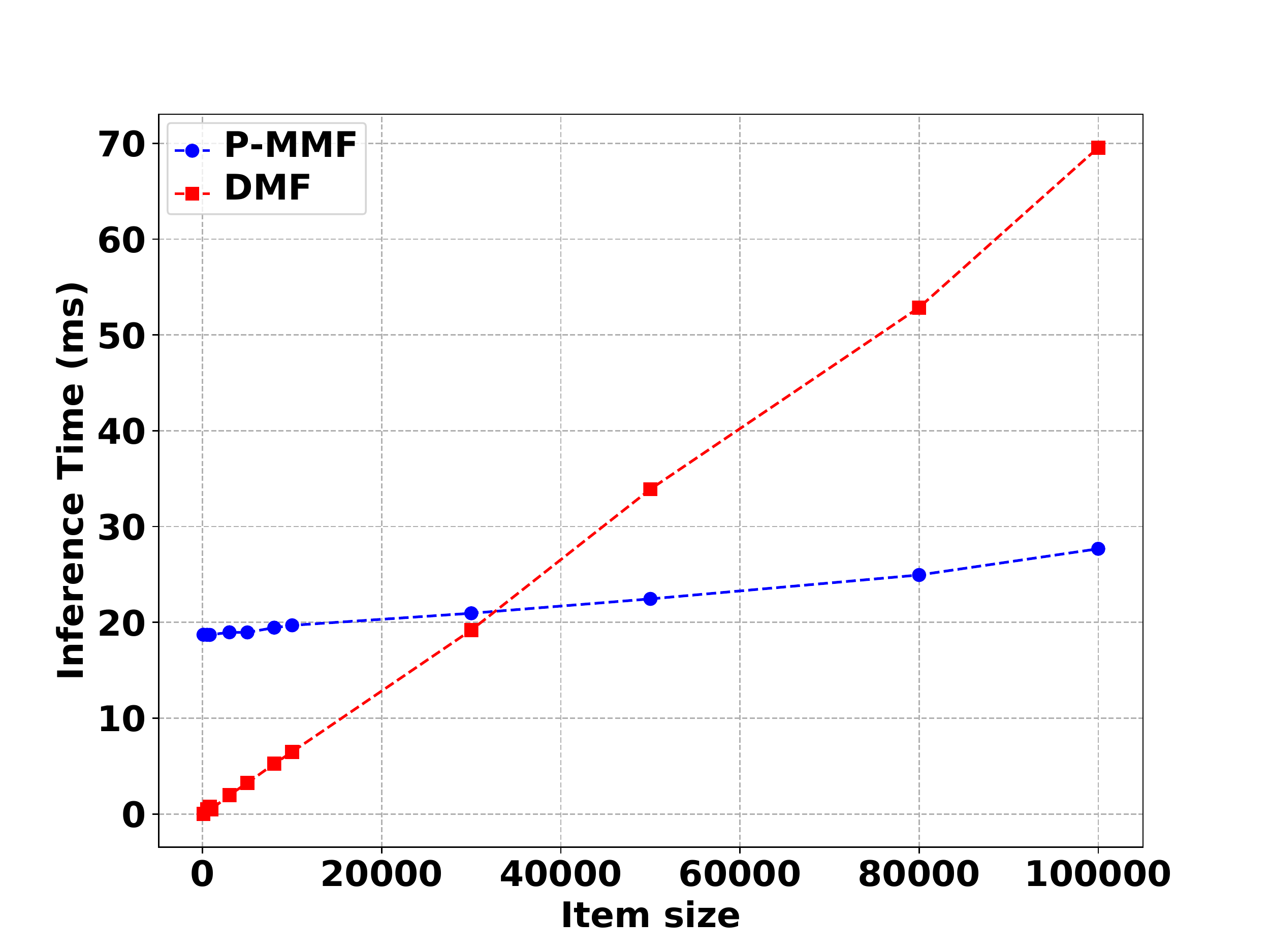}
    }
    \subfigure[GPU version]
    {
        \includegraphics[width=0.4\linewidth]{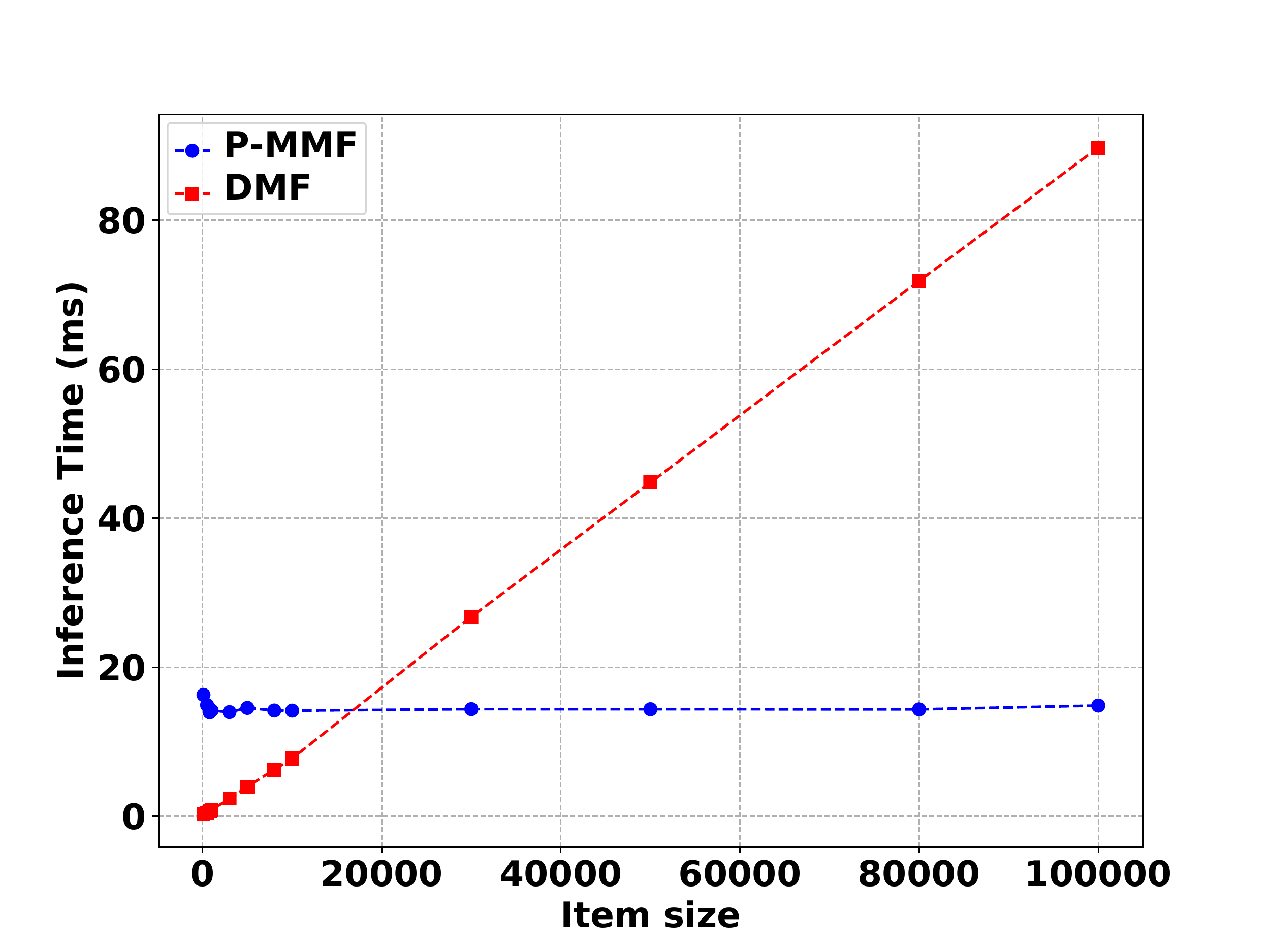}
    }

    \caption{Online inference time per user access under CPU and GPU implementations w.r.t. the number of total items. }
    \label{fig:inference_time}  
\end{figure}


\section{Conclusion}
This paper proposes a novel re-ranking model called P-MMF that aims to balance provider fairness and user preference.  Firstly, We formalize the task as a resource allocation problem regularized by a max-min fairness metric.  Then, to adapt the online scenarios in the recommendation, we proposed a momentum gradient descent method to conduct online learning for resource allocation in the dual space.  Theoretical analysis showed that the regret of P-MMF is bounded.  Extensive experimental results on four large-scale datasets demonstrated that P-MMF outperformed the baselines and Pareto dominated the state-of-the-art provider fair baselines.

\begin{acks}
This work was funded by the National Key R\&D Program of China (2019YFE0198200), the National Natural Science Foundation of China (61872338, 62006234, 61832017), Intelligent Social Governance Interdisciplinary Platform, Major Innovation \& Planning Interdisciplinary Platform for the ``Double-First Class'' Initiative and the Outstanding Innovative Talents Cultivation Funded Programs 2023 of Renmin University of China.
\end{acks}

\bibliographystyle{ACM-Reference-Format}
\balance
\bibliography{ref}

\newpage
\appendix
\section{Appendix}

\subsection{Proof of Theorem~\ref{theo:dual}}\label{app:dual_prove}
\begin{proof}
For max-min fairness, we have the regularizer as
$
    r(\mathbf{e}) = \min_{p\in\mathcal{P}} \left(\mathbf{e}_p/\bm{\gamma}_p\right),
$
we can easily proof that the exposure vector $\mathbf{e}$ can be represented as the dot-product between decision varible $\mathbf{x}_t$ and the item-provider adjacent matrix $\mathbf{A}$:
$
    \mathbf{e} = \sum_{t=1}^T\left(\mathbf{A}^{\top}\mathbf{x}_t\right).
$
Then we treat the $\mathbf{e}$ as the auxiliary variable, and the Equation~\eqref{eq:OPT} can be written as:
\begin{align*}
    W_{OPT} &= \max_{\mathbf{x}_t\in\mathcal{X},\mathbf{e}\leq \bm{\gamma}}\left[\sum_{t=1}^Tg(\mathbf{x}_t)/T + \lambda r(\mathbf{e})\right]\\
    s.t. \mathbf{e} &= \sum_{t=1}^T\left(\mathbf{A}^{\top}\mathbf{x}_t\right),
\end{align*}
where $\mathcal{X} = \{\mathbf{x}_t|\mathbf{x}_t \in {0,1} \land \sum_{i\in\mathcal{I}} \mathbf{x}_{ti} = K\}$. 
Then we move the constraints to the objective using a vector of Lagrange multipliers $\boldsymbol{\mu}\in\mathbb{R}^{|\mathcal{P}}|$:
\begin{align*}
    &W_{OPT} = \max_{\mathbf{x}_t\in\mathcal{X},\mathbf{e}\leq \bm{\gamma}}\min_{\boldsymbol{\mu}\in\mathcal{D}}\left[\sum_{t=1}^Tg(\mathbf{x}_t)/T + \lambda r(\mathbf{e}) - \boldsymbol{\mu}^{\top} \left(-\mathbf{e}+\sum_{t=1}^T\mathbf{A}^{\top}\mathbf{x}_t\right)\right]\\
    & \leq \min_{\boldsymbol{\mu}\in\mathcal{D}}\left[\max_{\mathbf{x}_t\in\mathcal{X}}\left[\sum_{t=1}^Tg(\mathbf{x}_t)/T - \boldsymbol{\mu}^{\top}\sum_{t=1}^T\mathbf{A}^{\top}\mathbf{x}_t\right] + \max_{\mathbf{e}\leq \bm{\gamma}}\left(\lambda r(\mathbf{e})-\boldsymbol{\mu}^{\top}\mathbf{e}\right)\right]\\
    &=\min_{\boldsymbol{\mu}\in\mathcal{D}}\left[f^*(\mathbf{A}\boldsymbol{\mu}) + \lambda r^{*}(-\boldsymbol{\mu})\right] = W_{Dual},
\end{align*}
where $\mathcal{D} = \boldsymbol{\mu}|r^*(-\boldsymbol{\mu})<\infty\}$ is the feasible region of dual variable $\boldsymbol{\mu}$. According to the Lemma 1 in the ~\citet{balseiro2021regularized}, we have $\mathcal{D}$ is convex and positive orthant is inside the recession cone of $\mathcal{D}$.

\end{proof}

\subsection{Proof of Theorem~\ref{theo:region}}\label{app:region_prove}
\begin{proof}
We let the variable $\mathbf{z}_p = (\mathbf{e}_p/\bm{\gamma}_p-1)$, we have:
\begin{align*}
    r^*(\boldsymbol{\mu}) &= \max_{\mathbf{e}\leq\bm{\gamma}}\left[\min\left(\mathbf{e}_p/\bm{\gamma}_p\right) + \boldsymbol{\mu}^{\top}\mathbf{\mathbf{e}}/\lambda\right]\\
    &= \boldsymbol{\mu}^{\top}\bm{\gamma} /\lambda +  1 + \max_{\mathbf{z}_p \leq 0}\left[\min(\mathbf{z}_p) + 1/\lambda \sum_{p\in\mathcal{P}}\boldsymbol{\mu}_p\bm{\gamma}_p\mathbf{z}_p\right]
\end{align*}
    
Let $s(\mathbf{z}) = \min_p \mathbf{z}_p$ and $\mathbf{v} = (\boldsymbol{\mu}\odot\bm{\gamma})/\lambda$, $\odot$ is the hadamard product. Then we define $s^*(\mathbf{v}) = \max_{\mathbf{z}\leq 0}\left(s(\mathbf{z}) + \mathbf{z}^T\mathbf{v}\right)$. We firstly show that if $\sum_{p\in\mathcal{S}}\mathbf{v}_p \ge -1, \forall \mathcal{S}\in\mathcal{P}_s$, then $s^*(\mathbf{v}) = 0$ and $\mathbf{z} = 0$ is the optimal solution, otherwise $s^*(\mathbf{v}) = \infty$.

We can equivalently write $\mathcal{D} = \{\mathbf{v}|\sum_{p\in\mathcal{S}}\mathbf{v}_p \ge -1, \forall \mathcal{S}\in\mathcal{P}_s\}$. We firstly show that $s^*(\mathbf{v}) = \infty$ for $\mathbf{v}\notin \mathcal{D}$. Suppose that there exists a subset $\mathcal{S}\in\mathcal{P}_s$ such that $\sum_{p\in\mathcal{S}} \mathbf{v}_p < -1$. For any $b > 1$, we can get a feasible solution:
\begin{align*}
\begin{split}
\mathbf{v}_p= \left \{
\begin{array}{ll}
   -b,                    & p\in \mathcal{S}\\
    0,                    & otherwise.
\end{array}
\right.
\end{split}
\end{align*}
Then, because such solution is feasible and $s(\mathbf{z}) = -b$, we obtain that 

\[s^*(\mathbf{v}) \ge s(\mathbf{z}) - b(\sum_{p\in\mathcal{S}}\mathbf{v}_p) = -b(\sum_{p\in\mathcal{S}}\mathbf{v}_p+1).
\]
Let $b\rightarrow\infty$, we have $s^*(\mathbf{v})\rightarrow\infty$.

Then we show that $s^*(\boldsymbol{\mu}) = 0$ for $\mathbf{v}\in\mathcal{D}$. Note that $\mathbf{z} = 0$ is feasible. Therefore, we have
\[
    s^*(\mathbf{v})\ge s^*(0) = 0.
\]
Then we have $\mathbf{z} \leq 0$ and without loss of generality, that the vector $\mathbf{z}$ is sorted in increasing order, i.e., $\mathbf{z}_1\leq \mathbf{z}_2, \cdots, \leq \mathbf{z}_{|\mathcal{P}|}$.
The objective value is
\begin{align*}
     s^*(\mathbf{v}) &= \mathbf{z}_1 + \sum_{j\in|\mathcal{P}|}\mathbf{z}_j\mathbf{v}_j \\
     &= \sum_{j=1}^{|\mathcal{P}|}\left(\mathbf{z}_j-\mathbf{z}_{j+1}\right)\left(1+\sum_{i=1}^{j}\mathbf{v}_j\right)\leq 0.
\end{align*}

Thus we can have $s^*(\boldsymbol{\mu}) = 0$ for $\mathbf{v}\in\mathcal{D}$ and we have 
\[
    r^*(-\boldsymbol{\mu}) = \boldsymbol{\mu}^{\top}\bm{\gamma}/\lambda + 1.
\]

\end{proof}

\subsection{Proof of Theorem~\ref{theo:regret}\label{app:regret_prove}}
\begin{proof}
Firstly, in practice, we normalize the user-item preference score $s_{u,i}$ to $[0,1]$. Therefore, $\sum_{t=1}^Tg(\mathbf{x}_t)/T\leq K$. In max-min regularizer $r(\mathbf{e})$. Let's  abbreviate its upper bound to $\bar{r}$. In practice, $\bar{r}\leq 1$ We have
\begin{equation}
    W_{OPT} \leq K + \lambda \bar{r}.
\end{equation}

We consider the stopping time $\tau$ of Algorithm~\ref{alg:P-MMF} as the first time the provider will have the maximum exposures, i.e.
\[
    \sum_{t=1}^{\tau}\mathbf{A}^{\top}\mathbf{x}_t \ge \bm{\gamma}.
\]
Note that is $\tau$ a random variable.

Similarly, followed the prove idea of~\citet{balseiro2021regularized}, First, we analysis the primal performance of the objective function. Second, we bound the complementary slackness term by the momentum gradient descent. Finally, We conclude by putting it to achieve the final regret bound.

\textbf{Primal performance proof}: Consider a time $t<\tau$, the recommender action will not violate the resource constraint. Therefore, we have:
\[
    g(\mathbf{x}_t)/T = g^*(\mathbf{A}\boldsymbol{\mu}_t) + \lambda {\boldsymbol{\mu}_t}^T\mathbf{A}^{\top}\mathbf{x}_t,
\]
and we have $\mathbf{e}_t = \argmax_{\mathbf{e}\leq \boldsymbol{\gamma}}\{r(\mathbf{e}) + \boldsymbol{\mu}^{\top}\mathbf{e}/\lambda\}$
\[
    r({\mathbf{e}_t}) = r^*(-\boldsymbol{\mu}) - \mathbf{\boldsymbol{\mu}_t}^T\mathbf{e}_t/\lambda.
\]

We make the expectations for the current time step $t$ for the primal functions:
\begin{align*}
    \mathbb{E}\left[g(\mathbf{x}_t)/T + \lambda  r({\mathbf{e}_t})\right] &= \mathbb{E}\left[g^*(\mathbf{A}\boldsymbol{\mu}_t) + \mathbf{\boldsymbol{\mu}_t}^T\mathbf{A}^{\top}\mathbf{x}_t + \lambda r^*(-\boldsymbol{\mu}) - \mathbf{\boldsymbol{\mu}_t}^T\mathbf{e}_t\right] \\
    &= W(\boldsymbol{\mu}_t) - \mathbf{E}\left[\boldsymbol{\mu}_t^T(-\mathbf{A}^{\top}\mathbf{x}_t + \mathbf{e}_t)\right].
\end{align*}

Consider the process $Z_t = \sum_{j=1}^T\boldsymbol{\mu}_j^t(-\mathbf{A}^{\top}\mathbf{x}_t + \mathbf{e}_t)-\mathbf{E}\left[\boldsymbol{\mu}_t^T(-\mathbf{A}^{\top}\mathbf{x}_t + \mathbf{e}_t)\right]$ is a martingale process. The Optional Stopping Theorem in martingale process~\cite{williams1991probability} implies that $\mathbb{E}\left[Z_{\tau}\right] = 0$. Consider the variable $w_t(\boldsymbol{\mu}_t) = \boldsymbol{\mu}_t^T(-\mathbf{A}^{\top}\mathbf{x}_t + \mathbf{e}_t)$, we have
\begin{align*}
    \mathbb{E}\left[\sum_{t=1}^{\tau}w_t(\boldsymbol{\mu}_t)\right] = \mathbb{E}\left[\sum_{t=1}^{\tau}\mathbb{E}\left[w_t(\boldsymbol{\mu}_t)\right]\right]
\end{align*}

Moreover, in MMF, the dual function $W_{Dual}$ is convex proofed in Theorem~\ref{theo:dual}, we have
\begin{equation}
\begin{aligned}
    \mathbb{E}\left[\sum_{t=1}^{\tau}g(\mathbf{x}_t)/T + \lambda r(\mathbf{e}_t)\right]  &= \mathbb{E}\left[\sum_{t=1}^{\tau}W_{Dual}(\boldsymbol{\mu}_t)\right] - \mathbb{E}\left[\sum_{t=1}^{\tau}w_t(\boldsymbol{\mu}_t)\right]\\
    &\leq \mathbb{E}\left[\tau W_{Dual}(\widetilde{\boldsymbol{\mu}_{\tau}})\right] - \mathbb{E}\left[\sum_{t=1}^{\tau}w_t(\boldsymbol{\mu}_t)\right],
\end{aligned}
\end{equation}
where $\widetilde{\boldsymbol{\mu}_{\tau}} =\sum_{t=1}^{\tau}\boldsymbol{\mu}_t/\tau$.

\textbf{Complementary slackness proof}
Then we aim to proof the complementary slackness $\sum_{t=1}^Tw_t(\boldsymbol{\mu}_t) - w_t(\boldsymbol{\mu})$ is bounded. Suppose there exists $G, s.t.$ the gradient norm is bounded $
\| \widetilde{\mathbf{g}}_t \| \leq G $. Then we have:
\begin{equation}\label{eq:regret_mu}
    \sum_{t=1}^{\tau}w_t(\boldsymbol{\mu}_t) - w_t(\boldsymbol{\mu}) \leq \frac{\lambda^2}{\eta} + \frac{G^2}{(1-\alpha)\sigma}\eta(\tau-1) + \frac{G^2}{2(1-\alpha)^2\sigma\eta},
\end{equation}
where the project function $\| \boldsymbol{\mu}-\boldsymbol{\mu}_t \|_{\boldsymbol{\gamma}}^2$ is $\sigma-$strongly convex.

Next we prove the inequality in Equation~\eqref{eq:regret_mu}. 
According to the Theorem 1 in ~\cite{momentum4online}, 
we have 
\[
    \| \mathbf{g}_t \|_2^2 = \|(1-\alpha)\sum_{s=1}^t\alpha^{t-s}(\widetilde{\mathbf{g}}_s)\|_2^2 \leq G^2,
\]
and 
\[
    \sum_{t=1}^{\tau}w_t(\boldsymbol{\mu}_t) - w_t(\boldsymbol{\mu}) \leq \frac{\|\boldsymbol{\mu}_t-\boldsymbol{\mu}_0\|_{\boldsymbol{\gamma}^2}^2}{\eta} + \frac{G^2}{(1-\alpha)\sigma}\eta(\tau-1) + \frac{G^2}{2(1-\alpha)^2\sigma\eta}, \forall \boldsymbol{\mu}.
\]
Assuming there exists $H>0$, s.t. $ \| \boldsymbol{\mu}_t-\boldsymbol{\mu}_0\|_{\boldsymbol{\gamma}^2}^2\leq H$. According to the Cauchy-Schwarz' inequality.
The results follows.
Let  $M=\frac{H}{\eta} + \frac{G^2}{(1-\alpha)\sigma}\eta(T-1) + \frac{G^2}{2(1-\alpha)^2\sigma\eta}$. 
We now choose a proper $\boldsymbol{\mu}$, s.t. the complementary stackness can be further bounded.

For $\boldsymbol{\mu} = \hat{\boldsymbol{\mu}} + \theta$, where $\theta\in\mathbb{R}^{|P|}$ is non-negative to be determined later and $\hat{\boldsymbol{\mu}} = \argmax_{\boldsymbol{\mu}}-\boldsymbol{\mu}^{\top}(\sum_{i=1}^T\mathbf{A}^{\top}\mathbf{x}_t)/\lambda$. According to the constraint $\mathbf{e} = \sum_{i=1}^T\mathbf{A}^{\top}\mathbf{x}_t$, we have that
\[
    \sum_{t=1}^T(r(\mathbf{e}_t) + \boldsymbol{\mu}^{\top}\mathbf{e}_t\lambda) \leq r^*(-\hat{\boldsymbol{\mu}}) = r(\sum_{i=1}^T\mathbf{A}^{\top}\mathbf{x}_t) + \hat{\boldsymbol{\mu}}^T(\sum_{i=1}^T\mathbf{A}^{\top}\mathbf{x}_t)/\lambda.
\]
Note that in proof of Theorem~\ref{theo:region}, the feasible region $\mathcal{D}$ is recession cone, therefore, $\boldsymbol{\mu}\in\mathcal{D}$.

Therefore, we have 
\begin{equation}
\begin{aligned}
   \sum_{t=1}^{\tau}w_t(\boldsymbol{\mu}_t)
   &= \sum_{t=1}^Tw_t(\hat{\boldsymbol{\mu}}) - \sum_{t=\tau+1}^Tw_t(\hat{\boldsymbol{\mu}})  + \sum_{t=1}^{\tau}w_t(\theta) + M.\end{aligned}
\end{equation}




\textbf{Put them together:} 
Under the max-min fair, we obtain that
\begin{equation}
    W_{OPT} = \frac{\tau}{T}W_{OPT} + \frac{T-\tau}{T}W_{OPT} \leq \tau W_{Dual}(\widetilde{\boldsymbol{\mu}_{\tau}}) + (T-\tau)(K+\lambda\bar{r}).
\end{equation}


Therefore, combining Eq.~(10,12,13) the regret $\text{Regret}(h)$ can be bounded as:
\begin{equation}
\begin{aligned}
    \text{Regret}(h) &= \mathbb{E}\left[W_{OPT}-W\right]\\
    &\leq \mathbb{E}\left[W_{OPT}-\sum_{t=1}^{\tau}\left(g(\mathbf{x}_t)/T - \lambda r(\mathbf{A}^{\top}\mathbf{x}_t/\boldsymbol{\gamma})\right)\right]\\
    &\leq \mathbb{E}\left[W_{OPT}-\tau W_{Dual}(\widetilde{\boldsymbol{\mu}_t}) + \sum_{t=1}^{\tau}w_t(\boldsymbol{\mu}_t) + \sum_{t=1}^T(\mathbf{e}_t-\mathbf{A}^{\top}\mathbf{x}_t)\right] \\
    &\leq \mathbb{E}\left[(T-\tau)(K+\lambda\bar{r})+\sum_{t=1}^Tw_t(\hat{\boldsymbol{\mu}}) + \sum_{t=1}^{\tau}w_t(\theta)\right] + M\\
    &\leq (T-\tau)(K+\lambda\bar{r}+\lambda K) + \sum_{t=1}^{\tau}w_t(\theta) + M.
\end{aligned}
\end{equation}

Let $C = K+\lambda\bar{r}+\lambda K$, then setting the $\theta = {C}{\min_p \boldsymbol{\gamma}_p}\mathbf{u}_p$, where $\mathbf{u}_p$ is the p-th unit vector. We have
\[
    \sum_{t=1}^{\tau}w_t(\theta) = C/(\min_p \boldsymbol{\gamma}_p)  - C(T-\tau).
\]
Then the $\text{Regret}(h)\leq M + C/(\min_p \boldsymbol{\gamma}_p)$, when we set $\eta = O(T^{-1/2})$, the Regret(h) is comparable with $O(T^{1/2})$. The result follows.


\end{proof}

\subsection{Algorithm for Min-reguarlizer}\label{app:min-regularizer}
In this section, we proposed a heuristic method for provider MMF online application, named Min-reguarlizer. It has a regularizer that measures the exposure gaps between the target provider and the worst-providers. The detailed algorithm is shown in Algorithm~\ref{alg:Min-Regularizer}. The notations are the same as P-MMF in Algorithm~\ref{alg:P-MMF}.

\begin{algorithm}[t]
    \caption{Online learning of Min-Regularizer}
	\label{alg:Min-Regularizer}
	\begin{algorithmic}[1]
	\REQUIRE User arriving set $\{u_i\}_{i=1}^N$, time-separate size $T$, ranking size $K$, user-item preference score $s_{u,i}, \forall u,i$, item-provider adjacent matrix $\mathbf{A}$, maximum resources $\bm{\gamma}$ and the trade-off coefficient $\lambda$.
	\ENSURE The decision variables $\{\mathbf{x}_i, i = 1,2,\cdots, N\}$
	\FOR{$n=1,\cdots,N/T$}
	    \STATE Initial dual solution $\boldsymbol{\mu}_1 = \mathbf{0}$. remain resources $\boldsymbol{\beta}_1 = \bm{\gamma}$ and momentum gradient $\mathbf{g}_0 = \mathbf{0}$.
	    \FOR{$t=1,\cdots,T$}
    	    \STATE Receive $u_{nT+t}$
    	    \STATE 
    	    \begin{align*}
            \begin{split}
            \mathbf{m}_p= \left \{
            \begin{array}{ll}
                0,                    & \boldsymbol{\beta}_{tp} > 0\\
                -\infty,                    & \textrm{otherwise}.
            \end{array}
            \right.
            \end{split}
            \end{align*}
\STATE $// ~~ \texttt{Make the recommendation:}$
    	    \STATE 
    	    $$
    	        \mathbf{x}_t = \argmax_{\mathbf{x}_t\in\mathcal{X}}\left[g(\mathbf{x}_t)-\lambda\left(\mathbf{A}(\mathbf{e} - (\min_p\mathbf{e}_p[1,1,\ldots,1]^{\top})+\mathbf{m})\right)^{\top}\mathbf{x}_t\right]
    	    $$
    	  \STATE  $// ~~ \texttt{Update the remaining resources:}$
    	    \STATE $$\boldsymbol{\beta}_{t+1} = \boldsymbol{\beta}_{t} - \mathbf{A}^{\top}\mathbf{x}_t$$
    	\ENDFOR
	\ENDFOR
	\end{algorithmic}
\end{algorithm}


\end{sloppy}
\end{document}